\documentclass[%
 aip,
 amsmath,amssymb,
 reprint,%
]{revtex4-1}

\usepackage{graphicx}
\usepackage{dcolumn}
\usepackage{bm}

\usepackage[utf8]{inputenc}
\usepackage[T1]{fontenc}
\usepackage{mathptmx}
\usepackage{etoolbox}
\usepackage{xcolor}
\usepackage{subfigure}
\usepackage[
    colorlinks=true,  
    linkcolor=blue,   
    citecolor=blue,  
    urlcolor=blue  
]{hyperref}

\makeatletter
\def\@email#1#2{%
 \endgroup
 \patchcmd{\titleblock@produce}
  {\frontmatter@RRAPformat}
  {\frontmatter@RRAPformat{\produce@RRAP{*#1\href{mailto:#2}{#2}}}\frontmatter@RRAPformat}
  {}{}
}%
\makeatother
\begin{document}

\preprint{AIP/123-QED}

\title[High-rate self-referenced continuous-variable quantum key distribution over high-loss free-space channel]{High-rate self-referenced continuous-variable quantum key distribution over high-loss free-space channel}
\author{Xiaojuan Liao}
 \affiliation{ 
State Key Laboratory of Advanced Optical Communication Systems and Networks, Center for Quantum Sensing and Information Processing, Shanghai Jiao Tong University, Shanghai 200240, China}%
\affiliation{ 
School of Electronic Information and Electrical Engineering, Shanghai Jiao Tong University, Shanghai 200240, China}%

\author{Yuehan Xu}%
\affiliation{ 
State Key Laboratory of Advanced Optical Communication Systems and Networks, Center for Quantum Sensing and Information Processing, Shanghai Jiao Tong University, Shanghai 200240, China}%
\affiliation{ 
School of Electronic Information and Electrical Engineering, Shanghai Jiao Tong University, Shanghai 200240, China}%

\author{Qijun Zhang}
\affiliation{ 
State Key Laboratory of Advanced Optical Communication Systems and Networks, Center for Quantum Sensing and Information Processing, Shanghai Jiao Tong University, Shanghai 200240, China}%
\affiliation{ 
School of Electronic Information and Electrical Engineering, Shanghai Jiao Tong University, Shanghai 200240, China}%

\author{Peng Huang}
\affiliation{ 
State Key Laboratory of Advanced Optical Communication Systems and Networks, Center for Quantum Sensing and Information Processing, Shanghai Jiao Tong University, Shanghai 200240, China}%
\affiliation{ 
School of Electronic Information and Electrical Engineering, Shanghai Jiao Tong University, Shanghai 200240, China}%
\affiliation{ 
Shanghai Research Center for Quantum Sciences, Shanghai 201315, China}%
\affiliation{ 
Hefei National Laboratory, Hefei 230088, China}%
\email{huang.peng@sjtu.edu.cn}

\author{Tao Wang}
\affiliation{ 
State Key Laboratory of Advanced Optical Communication Systems and Networks, Center for Quantum Sensing and Information Processing, Shanghai Jiao Tong University, Shanghai 200240, China}%
\affiliation{ 
School of Electronic Information and Electrical Engineering, Shanghai Jiao Tong University, Shanghai 200240, China}%
\affiliation{ 
Shanghai Research Center for Quantum Sciences, Shanghai 201315, China}%
\affiliation{ 
Hefei National Laboratory, Hefei 230088, China}%

\author{Kaizhi Wang}
\affiliation{ 
School of Electronic Information and Electrical Engineering, Shanghai Jiao Tong University, Shanghai 200240, China}%

\author{Guihua Zeng}
\affiliation{ 
State Key Laboratory of Advanced Optical Communication Systems and Networks, Center for Quantum Sensing and Information Processing, Shanghai Jiao Tong University, Shanghai 200240, China}%
\affiliation{ 
School of Electronic Information and Electrical Engineering, Shanghai Jiao Tong University, Shanghai 200240, China}%
\affiliation{ 
Shanghai Research Center for Quantum Sciences, Shanghai 201315, China}%
\affiliation{ 
Hefei National Laboratory, Hefei 230088, China}%
\affiliation{ 
Shanghai XunTai Quantech Co., Ltd, Shanghai, 200241, China}%
\email{ghzeng@sjtu.edu.cn}


\begin{abstract}
The advent of quantum computers has significantly challenged the security of traditional cryptographic systems, prompting a surge in research on quantum key distribution (QKD). Among various QKD approaches, continuous-variable QKD (CVQKD) offers superior resilience against background noise. However, the local local oscillator (LLO) CVQKD scheme faces substantial physical limitations in scenarios with high channel attenuation, and the large attenuation CVQKD remains unrealized. Bottleneck challenges include ensuring stable low-noise transmission and accurately estimating parameters under fluctuating channel conditions. In this paper, we introduce a continuous-time mode theory for high-precision estimation of time-varying parameters and design a free-space experimental system with a main quantum system and an auxiliary counterpart. We further develop advanced digital signal post-processing techniques for compensating time-varying frequency offset and phase noise under dynamic channel. Notably, the estimation of the time-varying free-space channel is achieved through the use of the auxiliary quantum system. Through experimental validation, we first demonstrate high-rate secure quantum key distribution over high-loss free-space channels. Specifically, we achieve asymptotic key rates of 76.366 kbps and 403.896 kbps in 25 dB attenuation free-space channels without turbulence and 21.5 dB average attenuation free-space channels with turbulence, respectively. Additionally, we confirm the feasibility of experiments on mildly turbulent atmospheric channels spanning at least 10.5 km using current equipments. Our scheme provides direct insight into constructing an integrated air-ground quantum communication network. 
\end{abstract}

\maketitle

%

\section{\label{section:1} Introduction}

With the rapid development of quantum computer technology, public key encryption systems based on computational complexity face significant challenges concerning the security of key distribution \cite{ref1,ref2,ref3,ref4}. The quantum key distribution (QKD) protocol, which relies on fundamental physical principles \cite{ref5}, allows remote users like Alice and Bob to establish unconditionally secure keys over an insecure quantum channel \cite{ref6,ref7,ref8,ref9,ref10}, effectively thwarting potential attacks \cite{ref11}. Theoretically, QKD protocols are classified into two main categories: discrete-variable QKD (DVQKD) and continuous-variable QKD (CVQKD). However, the secure key rate of QKD is limited by channel attenuation, which is considerably lower than that of classical communication systems. Consequently, finding solutions to achieve QKD in high-attenuation environments has become an urgent challenge that needs to be addressed.

In the field of fiber channels, DVQKD scheme based on the twin-field approach has successfully withstood channel attenuation of over 140 dB \cite{ref12}. Furthermore, it has the potential for secure key distribution even under a high attenuation of 157 dB \cite{ref13}. In comparison, DVQKD,  using the decoy-state BB84 protocol for free-space channels, has achieved secure key generation with a total attenuation of approximately 48 dB \cite{ref14,ref15,ref16}. CVQKD protocol offers the advantage of effectively suppressing background noise \cite{ref17,ref18,ref19,ref20}. This benefit arises from its coherent detection method and the inherent filtering characteristics of the local oscillator (LO) light \cite{ref7,ref21,ref22,ref23,ref24}. Specifically, the transmitted LO (TLO) CVQKD scheme has successfully achieved key distribution under attenuations of 30 dB \cite{ref25} and 32.45 dB \cite{ref26} in fiber channels. Meanwhile, the local LO (LLO) scheme \cite{ref27,ref28} has reached attenuation tolerances of 15.4 dB \cite{ref29} and 18 dB \cite{ref30}. Additionally, the feasibility of free-space CVQKD through atmospheric channels has been theoretically and experimentally verified \cite{ref31,ref32,ref33}. Among the various approaches, a Gaussian modulation-based scheme has only been implemented in a 490-meter atmospheric channel \cite{ref34}. An experiment based on passive-state-preparation generated a secure key in a simulated turbulent free-space channel with an attenuation of 15 dB \cite{ref20}. Table~\ref{table1} compares the performance of these CVQKD experiments in recent years. However, not all schemes can simultaneously ensure high code rate transmission in large-attenuation free-space channels.

\begin{table*}
\caption{\label{table1}Performance comparison of CVQKD experiments in recent years}
\begin{ruledtabular}
\begin{tabular}{cccccccc}
Reference & Channel & Laser Source & LO & Loss & Repetition Rate & Secret Key Rate & T Estimation \\ \hline
Ref.~\onlinecite{ref26} & Fiber & CW & TLO & 32.45 dB & 5 MHz & 6.214 bps & fixed \\
Ref.~\onlinecite{ref30} & Fiber & CW & LLO & 18 dB & 1 GHz & 0.51 Mbps & fixed \\
Ref.~\onlinecite{ref34} & Free space & CW & TLO & 1.87 dB & 10 kHz & 0.152 kbps & average \\
Ref.~\onlinecite{ref20} & Free space & ASE & TLO & 15 dB & 350 MHz & 1.015 Mbps & average \\
Current work & Free space & CW & LLO & 21.5 dB & 1 GHz & 403.896 kbps & Real-time \\
\end{tabular}
\end{ruledtabular}
\end{table*}

In previous studies, scholars have designed LLO fiber systems \cite{ref29,ref30,ref35} aiming to solve the frequency and phase-locking problems between different lasers. Technologies such as phase compensation \cite{ref36}, dynamic polarization control \cite{ref37}, and excess noise suppression \cite{ref38} have laid a solid foundation for the realization of free-space LLO-CVQKD schemes. However, when applying the LLO-CVQKD system to high-attenuation environments, significant challenges, such as excess noise suppression and high stability control, still need to be addressed. Meanwhile, free-space systems must also overcome the problem of rapid changes in channel transmittance. In past studies, the evaluation of channel transmittance in the process of free-space QKD was often limited to estimating the average transmittance of signal frames. However, this fixed estimation method cannot capture the dynamic characteristics of channel transmittance changes over time. It may cover up instantaneous fluctuations in channel transmission, thereby misleading the accurate assessment of system performance.

In this paper, we experimentally verify secure key distribution with high attenuation and high rate in a free-space LLO-CVQKD system based on the estimation of time-varying parameters. In particular, we establish a continuous-mode theoretical model that accounts for time-varying frequency offset, phase noise, and modulation imbalance. Additionally, digital signal processing (DSP) algorithms are obtained to achieve the mode-matching and time-varying transmittance estimation and time-varying compensation at the receiver, enabling high-precision parameters evaluation and excess noise suppression, even under high-speed and significant, time-varying attenuation conditions. Specifically, our work presented a ground-breaking innovation in the form of free-space CVQKD with a 1 GHz repetition frequency over 25 dB channel attenuation under laboratory conditions without turbulence, achieving an asymptotic key rate of 76.366 kbps. While under the turbulence channel with an average attenuation of 21.5 dB, the system could generate an asymptotic key rate of 403.896 kbps, all while enabling real-time estimation of the free-space channel transmittance during the key distribution process. This research provides a direct reference for developing an integrated ground-to-space quantum communication network.

\section{\label{section:2} Time-varying Estimation Model Using Continuous Temporal Modes}

\subsection{\label{section:2.1} Time-varying factors in the LLO system}

Quadrature modulation is often used to prepare coherent states in CVQKD systems \cite{ref8}. In practice, quadrature modulation can be implemented using in-phase and quadrature modulator (IQM), and the IQM can be utterly equivalent to a system consisting of beam splitters (BS) and a double nested amplitude modulators (AM) and a phase modulator (PM) \cite{ref26,ref39,ref40,ref41}.

\begin{figure}
\includegraphics[width=0.48\textwidth]{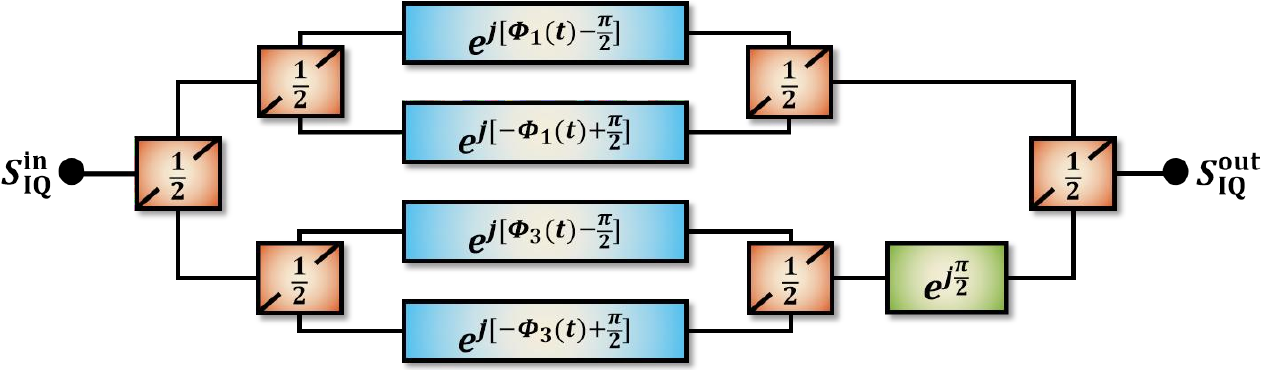}
\caption{\label{fig:2.1-1} The ideal IQ modulator model. The orange rectangle represents the BS. The two blue rectangles at the top and the two blue rectangles at the bottom represent a pair of nested AMs. The green rectangle represents PM.}
\end{figure}

In an ideal balanced scenario, where the reflection coefficient of each BS is 0.5, the bias phase of each pair of AM is $+\pi/2$ and $-\pi/2$, respectively. The orthogonal bias point of the PM is located at $\pi/2$, as illustrated in Fig.~\ref{fig:2.1-1}. The relationship between the input and output of the IQM can be expressed as follows:
\begin{equation}
    \label{eq2.1-1}
    S_{\rm{IQ}}^{\rm{out}}=\left[sin\left(\phi_1(t)\right)+j\cdot sin\left(\phi_3(t)\right)\right]\cdot S_{\rm{IQ}}^{\rm{in}}.
\end{equation}

An imperfect BS can lead to a power imbalance between the I and Q paths in non-ideal cases. This alteration in the orthogonal bias point may result in a loss of orthogonality between the I and Q paths. Consequently, the modulation imbalance affects the constellation, causing it to stretch and skew along the diagonal \cite{ref41}. To account for the effects of time-varying modulation imbalance, we can rewrite the input-output relationship of the IQM.
\begin{equation}
    \label{eq2.1-2}
    S_{\rm{IQ}}^{\rm{out}}=\left[a(t)sin\left(\phi_1(t)\right)+j\cdot b(t)e^{j\theta_m(t)}sin\left(\phi_3(t)\right)\right]\cdot S_{\rm{IQ}}^{\rm{in}},
\end{equation}
\noindent in which, $a(t)$ and $b(t)$ reflects the power imbalance, $\theta_m(t)$ represents the change in the orthogonal bias point.

In an imperfect modulation scenario, the signal constellation doesn't align with the ideal version. If Alice and Bob mistakenly assume that the constellation diagram remains flawless, the precision in estimating channel parameters will deteriorate.

\begin{figure*}
\includegraphics[width=1\textwidth]{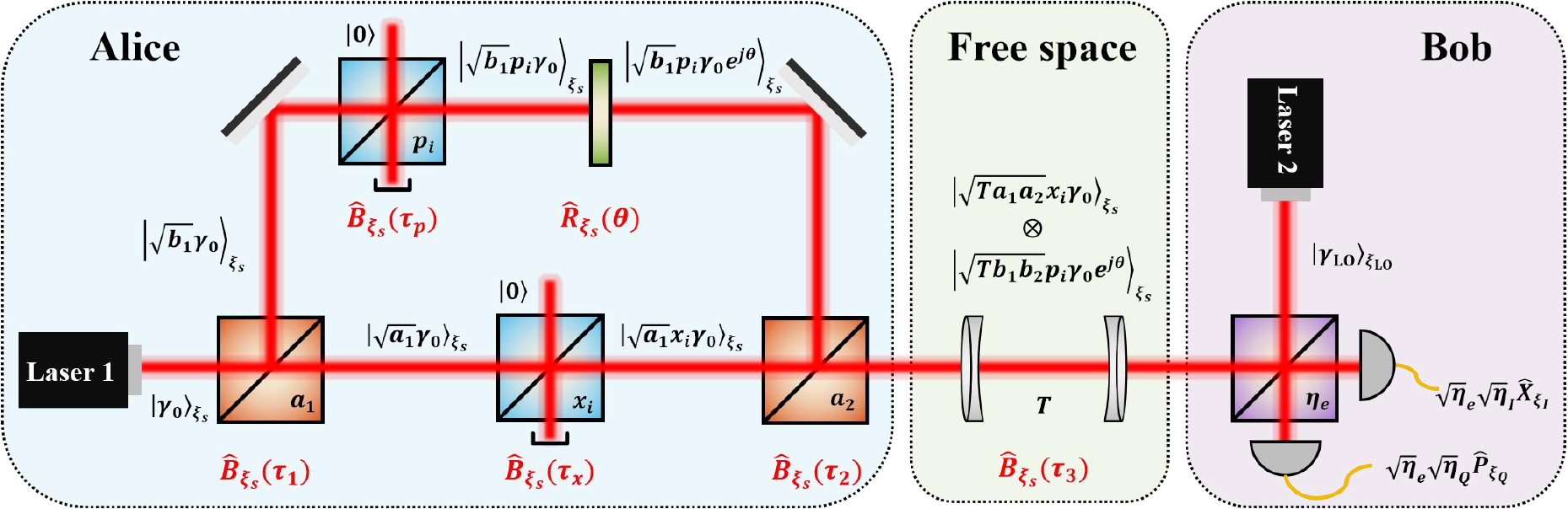}
\caption{\label{fig:2.2-1}LLO-CVQKD theoretical model based on temporal modes of continuous-mode states. The orange rectangle represents the optical BS operator $\hat{B}_{\xi_s}()$, the blue rectangle represents the BS operator that performs the intensity modulation, and the green rectangle represents the phase rotation operator $\hat{R}_{\xi_s}()$. The effect of a free-space channel with transmittance T is considered equivalent to that of the BS operator $\hat{B}_{\xi_s}()$. The purple rectangle represents a heterodyne detector with a detection efficiency of $\eta_e$.}
\end{figure*}

Accurate transmittance estimation is essential for improving the secure key generation rate and system security. Although traditional average transmittance estimation is straightforward \cite{ref20,ref34}, it fails to effectively account for the channel's time-varying nature. In real free-space environments, the propagation of optical signals is influenced by weather changes, optical obstacles, and atmospheric turbulence. Therefore, employing a time-varying transmittance estimation approach allows for a more precise representation of the channel's dynamic characteristics, ultimately enhancing the performance of the CVQKD system.

In conclusion, we require a set of high-precision theoretical models for parameter estimation to compensate for performance deterioration caused by time-varying factors. This will help achieve excess noise suppression under significant attenuation.

\subsection{\label{section:2.2} Time-varying parameters estimation model}

Based on the fundamental definition of temporal modes of continuous-mode states \cite{ref42,ref43,ref44,ref45,ref46,ref47}, along with the displacement operator, phase rotation operator, and beam splitter model in continuous mode, detailed information can be found in Appendix~\ref{Appendix:A1} and Appendix~\ref{Appendix:A2}, it is possible to obtain the theoretical estimation of time-varying parameters in LLO-CVQKD.

Fig.~\ref{fig:2.2-1} illustrates the continuous mode theoretical model of the LLO-CVQKD system. At the Alice, the photon-wavepacket coherent state first passes through the IQ modulator. Specifically, it is subjected to an optical BS operator $\hat{B}_{\xi_s}(\tau_1)$ with a parameter $\tau_1=cos^{-1}a_1$, which divides it into two photon-wavepacket coherent states along the I and Q paths. The I and Q components are modulated separately, by different optical BS operators with parameters $\tau_x=cos^{-1}x_i$ and $\tau_p=cos^{-1}p_i$ respectively. Then, the Q path is orthogonal to the I path under the action of the phase rotation operator. Finally, the coherent states are combined again by the BS operator $\hat{B}_{\xi_s}(\tau_2)$ with the parameter $\tau_2=cos^{-1}a_2$. The free-space channel is modeled as an optical BS operator. After the action of the operation with the parameter $\tau_3=cos^{-1}T$, the coherent state is ultimately detected by Bob through heterodyne detection with detection efficiency $\sqrt{\eta_e}$. We will describe the derivation process involving modulation, transmission, and detection.

Alice randomly generates two arrays
\begin{equation}
\label{eq2.2-2}
\begin{split}
    &\left\{x_q(i)\right\}\in\left\{x_1,x_2,\ldots,x_n\right\},\\
    &\left\{p_q(i)\right\}\in\left\{p_1,p_2,\ldots,p_n\right\}.
\end{split}
\end{equation}

Both arrays have a length of  \(n\)  and are characterized by a Gaussian distribution with a mean of zero. Following Nyquist's first criterion, the baseband shaping filter used at the transmitter and the matched filter at the receiver are both square root raised cosine (RRC) roll-off filters, which are standard in practical communication systems. First, we must obtain the RRC FIR filter coefficients with a roll-off factor of \(\alpha\). This filter is truncated to span a certain number of symbols, with \(sps\) samples representing each symbol. The RCOSDESIGN function designs a symmetric filter, and its order is determined by \(sps \times span\). Eq.~(\ref{eq2.2-1}) gives the time-domain impulse response of this RRC filter, where \(n \in \left[1, sps \times span + 1\right]\). Before applying baseband shaping, we need to interpolate the signals  \(\left\{x_q(i)\right\}\) and \(\left\{p_q(i)\right\}\) by a zero-padding factor. This will allow for proper sampling of the signals.
\begin{equation}
    \label{eq2.2-3}
    \begin{split}
        &{\left\{x_q^{\rm{up}}(i)\right\}\in\{x_1,0,0,\ldots,x_2,0,0,\ldots,x_n,0,0,\ldots\}},\\
        &{\left\{p_q^{\rm{up}}(i)\right\}\in\{p_1,0,0,\ldots,p_2,0,0,\ldots,p_n,0,0,\ldots\}}.
    \end{split}
\end{equation}

Both lengths are \( L_{\rm{up}} = n \times sps \). Signal filtering using the RRC filter involves a convolution operation between the filter taps and the upsampled signals \(\left\{x_q^{\rm{up}}(i)\right\}\) and \(\left\{p_q^{\rm{up}}(i)\right\}\). The modulation is assumed to start at time \( t_0 \), and the baseband modulation rate is \( F_m \). At time \( t \in \left(t_0, t_0 + N/F_m\right) \), the I and Q components are represented as 
$\left\{x_q^m(t)\right\} \in \left\{x_q^{\rm{up}}\right\} \ast \left\{h_{\rm{rrc}}\right\}$ and 
$\left\{p_q^m(t)\right\} \in \left\{p_q^{\rm{up}}\right\} \ast \{h_{\rm{rrc}}\}$,
which correspond to the encoding quantum signals. The lengths of both are given by 
$N = L_h + L_{\rm{up}} - 1$.

\begin{widetext}
\begin{equation}
h_{\rm{rrc}}(n)=
    \begin{cases}
        -\frac{1}{\pi\cdot sps}\left[\pi(\alpha-1)-4\alpha\right],&n=\frac{span}{2}\\
            \frac{1}{2\pi\cdot sps}\left[\pi\left(\alpha
        +1\right)sin\frac{\pi(\alpha+1)}{4\alpha}
        -4\alpha sin\frac{\pi(\alpha-1)}{4\alpha}
         +\pi(\alpha-1)cos\frac{\pi(\alpha-1)}{4\alpha}\right],&n<span\quad and\quad \left||4\alpha(n-\frac{span}{2})|-1\right|<0\\
           -\frac{4\alpha}{\pi\cdot sps\left[(4\alpha \left(n-\frac{span}{2}\right)^2-1\right]}
           \left\{
           cos\left[\pi(1+\alpha)\left(n-\frac{span}{2}\right)\right]
            +\frac{sin[\pi(1-\alpha)n]}{4\pi \left(n-\frac{span}{2}\right)}\right\},&n<span\quad and \quad \big||4\alpha(n-\frac{span}{2})|-1\big|\geq0\\
            0,&otherwise
    \end{cases}. 
    \label{eq2.2-1}
\end{equation}
\end{widetext}

For the pilot signal, we have \( \left\{x_p(i)\right\} \in \left\{A_p, A_p, \ldots, A_p\right\} \) and \( \left\{p_p(i)\right\} \in \left\{A_p, A_p, \ldots, A_p\right\} \), both of which have a length of \( n \), where \( A_p \) represents the amplitude of the pilot. The pilot signal undergoes the same RRC filtering to produce the encoding pilot signal \( A_p^m(t) \). 

Additionally, \( \left\{A_p^m(t) \cos(\omega_p t)\right\} \) and \( \left\{A_p^m(t) \sin(\omega_p t)\right\} \) are modulated by Alice using carrier modulation, where \( \omega_p \) is the carrier frequency of the pilot signal. The final frequency-division multiplexed (FDM) signal modulated by Alice is
\begin{equation}
    \label{eq2.2-4}
    \begin{split}
        &\left\{x_q^m(t) + A_p^m(t) \cos(\omega_p t)\right\},\\
        &\left\{p_q^m(t) + A_p^m(t) \sin(\omega_p t)\right\}.
    \end{split}
\end{equation}

Laser 1 at Alice generates the photon-wavepacket coherent state as \( \left| \gamma_0 \right\rangle_{\xi_s} \). Here, \( \xi_s(t) = \xi^0_s(t) e^{-j \omega_s(t) t} \), with \( \omega_s(t) \) being the optical carrier of the signal. The quantum and pilot signals are loaded onto this same envelope via FDM. The process is equivalent to dividing the photons into two parts to modulate the quantum and pilot signals.

In actual experiments, the BS in the IQM setup is not ideally balanced, and the modulated signals are not perfectly orthogonal. After the coherent state \( \left| \gamma_0 \right\rangle_{\xi_s} \) enters the IQM, it first passes through an imbalanced BS, dividing it into two paths: the I-path and the Q-path. The square root transmittance of this imbalanced BS is denoted as \( a_1(t) \), while the corresponding reflectance is given by \( b^2_{1}(t) = 1 - a^2_{1}(t) \). The process affecting \( \left| \gamma_0 \right\rangle_{\xi_s} \) is as follows:
\begin{equation}
    \label{eq2.2-5}
    \begin{split}
        \hat{B}_{\xi_s}\left[cos^{-1}a_1(t)\right]\left|0\right\rangle\otimes\left|\gamma_0\right\rangle_{\xi_s}
        =\left|a_1(t)\gamma_0\right\rangle_{\xi_s}\otimes \left|b_1(t)\gamma_0\right\rangle_{\xi_s}.
    \end{split}
\end{equation} 

Intensity modulation can generate a new coherent state related to the modulated signal using the optical BS operator. In the following derivation, the reflected light part is ignored. For the I-path signal,
\begin{equation}
    \label{eq2.2-6}
    \begin{split}
    &\left|I_{\rm{AM}}\right\rangle_{\xi_{s}}\\
    =&\hat{B}_{\xi_{s}}\left[cos^{-1}\left(x_q^m(t)+A_p^m(t)cos(\omega_pt)\right)\right]\left|0\right\rangle\otimes\left|a_1(t)\gamma_0\right\rangle_{\xi_s}\\
    =&\left|\left[x_q^m(t)+A_p^m(t)cos(\omega_pt)\right]a_1(t)\gamma_0\right\rangle_{\xi_{s}}.\\
    \end{split}
\end{equation}

Similarly, we can also obtain the Q-path signal: 
\begin{equation}
    \label{eq2.2-7}
    \left|Q_{\rm{AM}}\right\rangle_{\xi_{s}}=\left|\left[p_q^m(t)+A_p^m(t)sin(\omega_pt)\right]b_1(t)\gamma_0\right\rangle_{\xi_{s}}.
\end{equation}

Under ideal conditions, once the Q-path signal passes through the phase modulator, it should be orthogonal to the I-path signal. However, the PM does not accurately achieve a phase shift of \(\frac{\pi}{2}\). Let's assume the extra rotation angle of the Q-path is represented by \(\theta_m(t)\), which varies with time. By applying the phase rotation operator \(\hat{R}_{\xi_s} \left[ \frac{\pi}{2} + \theta_m(t) \right]\) to the Q-path signal, we can account for this time-varying phase shift.
\begin{equation}
    \label{eq2.2-8}
    \begin{split}
    \left|Q_{\rm{PM}}\right\rangle_{\xi_s}=&\hat{R}_{\xi_s}\left[\frac{\pi}{2}+\theta_m(t)\right]\left|Q_{\rm{AM}}\right\rangle_{\xi_{s}}\\
    =&\left|j\left[p_q^m(t)+A_p^m(t)sin(\omega_pt)\right]b_1(t)e^{j\theta_m(t)}\gamma_0\right\rangle_{\xi_{s}}.\\
    \end{split}
\end{equation}

Finally, after passing through another imbalanced BS with a square root transmittance of \( a_2(t) \) and a reflectance of \( b_2(t) \), apply the BS operators \( \hat{B}_{\xi_{s}}\left[\cos^{-1}a_2(t)\right] \) to the coherent states in both the I and Q paths. This allows the signals to be directed together to complete the combined output.
\begin{equation}
    \label{eq2.2-9}
    \begin{split}
        &\hat{B}_{\xi_s}\left[cos^{-1}a_2(t)\right]\left|I_{\rm{AM}}\right\rangle_{\xi_s}\otimes
        \left|Q_{\rm{PM}}\right\rangle_{\xi_s}\\
=& \left|x_q^m(t)a(t)\gamma_0+jp_q^m(t)b(t)e^{j\theta_m(t)}\gamma_0\right\rangle_{\xi_{s}}
         \\
\otimes& 
\left|A_p^m(t)cos(\omega_pt)a(t)\gamma_0+jA_p^m(t)sin(\omega_pt)b(t)e^{j\theta_m(t)}\gamma_0\right\rangle_{\xi_s},
    \end{split}
\end{equation}
in which, $a(t)=a_1(t)a_2(t)$ and $b(t)=-b_1(t)b_2(t)$.

Ten percent of the coherent state is directed through a BS and enters automatic bias control (ABC) for bias control. The remaining ninety percent is routed to a variable optical attenuator (VOA) for power adjustment. After proper calibration, Alice produces the photon-wavepacket coherent state output:
\begin{equation}
    \label{eq2.2-10}
    \begin{split}
       & \left|x_q^m(t)+jp_q^m(t)d(t)e^{j\theta_m(t)}\right\rangle_{\xi_{s}}
         \\
        \otimes& 
        \left|A_p^m(t)cos(\omega_pt)+jA_p^m(t)sin(\omega_pt)d(t)e^{j\theta_m(t)}\right\rangle_{\xi_s}.
    \end{split}
\end{equation}

Here, the first term represents the main quantum system, and the second represents the auxiliary one. $d(t)=b(t)/a(t)$, which represents the ratio of the imbalance between the I and Q paths. Alice's modulation variance is 
\begin{equation}
\label{eq2.2-11}
    V_A=2{\left|x_q(i)+jp_q(i)\right|}^2.
\end{equation}

Afterward, the photon-wavepacket coherent state enters the free-space channel. During this process, it undergoes phase rotation, power attenuation, and changes in polarization. The phase rotation caused by the channel is represented by \({\theta}_s(t)\). The transmittance of the free-space channel, \(T(t)\), varies with time. The changes in polarization will be addressed later. Utilizing the phase rotation operator \(\hat{R}_{\xi_s} \left[\theta_s(t) \right]\) and the beam splitter model \( \hat{B}_{\xi_s}\left[cos^{-1}T(t)\right] \), we can describe the photon-wavepacket coherent state as it reaches Bob.
\begin{equation}
    \label{eq2.2-12}
    \begin{split}
    &\left|\gamma_{\rm{quan}}\right\rangle_{\xi_{\rm{s1}}}\otimes \left|\gamma_{\rm{pilot}}\right\rangle_{\xi_{\rm{s1}}}\\
    =&\left|\sqrt{T(t)}\left[x_q^m(t)+j \cdot p_q^m(t)d(t)e^{j\theta_m(t)}\right]\right\rangle_{\xi_{\rm{s1}}}
         \\
    \otimes& 
    \left|\sqrt{T(t)}A_p^m(t)\left[cos(\omega_pt)+j\cdot sin(\omega_pt)d(t)e^{j\theta_m(t)}\right]\right\rangle_{\xi_{\rm{s1}}}.\\
    \end{split}
\end{equation}

In the formula, $\xi_{\rm{s1}}(t)=\xi_s(t)e^{j\theta_s(t)}=\xi_s^0(t)e^{j(-\omega_s(t)+\theta_s(t))}$. When the photon-wavepacket coherent state arrives at Bob, it first passes through the polarization controller for polarization correction. Assuming the polarization controller is ideal, polarization leakage is not considered. Bob's Laser 2 generates another photon-wavepacket coherent state, represented as \({\left|{\gamma}_{\rm{LO}}\right\rangle}_{{\xi}_{\rm{LO}}}\), where \({\xi}_{\rm{LO}}(t)={\xi}_{\rm{LO}}^0 e^{j\left(-\omega_{\rm{LO}}(t)t+\theta_{\rm{LO}}(t)\right)}\). \({\omega}_{\rm{LO}}(t)\) represents the optical carrier frequency of the LO, and \({\theta}_{\rm{LO}}(t)\) indicates the phase difference between Laser 1 and Laser 2. Using heterodyne detection, Bob inputs these states into coherent receivers (ICR) with a detection efficiency of $\eta_e$ for two orthogonal measurement modes. There are two orthogonal TMs with different but correlated displacement parameters. The corresponding output data \(\hat{I}_N^{\rm{SNU}}\) in shot noise units (SNU), with a block size of \(N\), can be obtained from the photocurrent flux operator. Once the photon-wavepacket coherent state Eq.~(\ref{eq2.2-12}) enters Bob's ICR for heterodyne detection, we can rewrite the measurement result through the Gram-Schmidt process, which can be given by
\begin{equation}
    \label{eq2.2-13}
    \begin{split}
 \hat{I}_{\xi_{\rm{dsp}}}^{\rm{SNU}} &=\hat{X}_{\xi_{\rm{dsp}}}+j\cdot\hat{P}_{\xi_{\rm{dsp}}} =\sqrt{\eta_e}\sqrt{\eta}\left(\hat{X}_{\xi_{\rm{s1}}}+j\cdot\hat{P}_{\xi_{\rm{s1}}}\right).\\
\end{split}
\end{equation}

Here, $\sqrt{\eta}$ can be considered as the mode-matching coefficient. This calculation's specific procedures and detailed derivations can be found in Appendix~\ref{Appendix:A3}.

First, let the DSP function operate like a filter $f_{\rm{dsp1}}^m=H_p^m$, where $H_p^m$ is the discrete impulse response function of the band-pass filter at $m$-th data. Thus, we can separate the pilot signal $\left|\gamma_{\rm{pilot}}\right\rangle_{\xi_{\rm{s1}}}$, which matches the frequency band of the band-pass filter $H_p^m$.
\begin{equation}
    \label{eq2.2-14}
    \begin{split}
        \sqrt{\eta_1}=&\int_{t_0}^{t_0+N/F_m}dt\frac{1}{\sigma_{\rm{cal}}^*}\xi_{\rm{LO}}^*(t){G_{\rm{dsp1}}^{N^*}}(t)\xi_{\rm{s1}}(t)\\
        =&\int_{t_0}^{t_0+N/F_m}dt\frac{1}{\sigma_{\rm{cal}}^*}G_{\rm{dsp1}}^{N^*}\xi_{\rm{LO}}^{0^*}(t)\xi_s^0(t)\\
        &exp\{j\left[\omega_{\rm{LO}}(t)t-\omega_s(t)t-\theta_{\rm{LO}}(t)+\theta_s(t)\right]\}.
    \end{split}
\end{equation}

Write $\alpha(t)=\omega_{\rm{LO}}(t)t-\omega_s(t)t-\theta_{\rm{LO}}(t)+\theta_s(t)$. Thus, we can get the first-order moment $\mu_1$ (details are in Appendix~\ref{Appendix:A4}):
\begin{equation}
    \label{eq2.2-15}
    \begin{split}
        \mu_1 = &\left\langle\gamma_{\rm{pilot}}\right|\hat{I}_{\xi_{\rm{dsp}}}^{\rm{SNU}}\left|\gamma_{\rm{pilot}}\right\rangle_{\xi_{\rm{s1}}}\\ =&\left\langle\gamma_{\rm{pilot}}\right|\sqrt{\eta_e\eta_1}(\hat{X}_{\xi_{\rm{s1}}}+j\cdot\hat{P}_{\xi_{\rm{s1}}})\left|\gamma_{\rm{pilot}}\right\rangle_{\xi_{\rm{s1}}}\\
        =&\frac{1}{2}\sqrt{T(t)\eta_e}A_p^m(t)    \left[d(t)e^{j\theta_m(t)}+1\right]e^{j[\omega_pt+\alpha(t)]}\\
        +&\frac{1}{2}\sqrt{T(t)\eta_e}A_p^m(t)
        \left[1-d(t)e^{j\theta_m(t)}\right]e^{j[-\omega_pt+\alpha(t)]}.
    \end{split}
\end{equation}

Next, we will implement time-varying frequency offset recovery and phase noise recovery. We start by performing high-pass filtering on \(\mu_1\) to obtain the part containing $e^{j[\omega_pt+\alpha(t)]}$, followed by amplitude normalization to achieve an exponential part with an amplitude of 1. Since \(e^{j\omega_pt}\) is known, we can extract the components affected by time-varying frequency offset and phase noise. This set of operations represents $f^{N}_{\rm{dsp2}}$. Thus, we can get a new first-order moment $\mu_2$.
\begin{equation}
    \label{eq2.2-16}
    \mu_2= e^{j[\omega_{\rm{LO}}(t)t - \omega_s(t)t - \theta_{\rm{LO}}(t) + \theta_s(t)]}.
\end{equation}

Now we can get the DSP algorithm $f^{N}_{\rm{dsp3}} = \mu_2$. This expression indicates that the signal is being rotated, enabling us to compensate for both time-varying frequency offset and phase noise. After compensation, the new mode-matching coefficient and first-order moment is
\begin{equation}
    \label{eq2.2-17}
    \begin{split}
        \sqrt{\eta_3}
        =&\int_{t_0}^{t_0+N/F_m}dt\frac{1}{\sigma_{\rm{cal}}^*}G_{\rm{dsp3}}^{N^*}\xi_{\rm{LO}}^*(t)\xi_{\rm{s1}}(t)\\
        =&\int_{t_0}^{t_0+N/F_m}dt\frac{1}{\sigma_{\rm{cal}}^*}\xi_{\rm{LO}}^{0^*}(t)\xi_{s}^0(t),
    \end{split}
\end{equation}
\begin{equation}
    \label{eq2.2-18}
    \begin{split}
        \mu_3 =&\left\langle\gamma_{\rm{pilot}}\right|\sqrt{\eta_e\eta_3}(\hat{X}_{\xi_{\rm{s1}}}+j\cdot\hat{P}_{\xi_{\rm{s1}}})\left|\gamma_{\rm{pilot}}\right\rangle_{\xi_{\rm{s1}}}\\
        =&\sqrt{T(t)\eta_e}A_p^m(t)\left[cos(\omega_pt)+j\cdot d(t)e^{j\theta_m(t)}sin(\omega_pt)\right]\\
        =&\sqrt{T(t)\eta_e}A_p^m(t)\left[cos(\omega_pt)-d(t)sin(\theta_mt)sin(\omega_pt)\right]\\
        +j&\sqrt{T(t)\eta_e}A_p^m(t)\left[sin(\omega_pt)+(d(t)cos(\theta_mt)-1)sin(\omega_pt)\right].
    \end{split}
\end{equation}

$\mu^{\rm{CD}}_3$ is obtained by coherent demodulation (CD) of $\mu_3$ and the effect of carrier is removed.
\begin{equation}
    \label{eq2.2-19}
    \begin{split}
        \mu^{\rm{CD}}_3 
       =&\sqrt{T(t)\eta_e}A_p^m(t)\left[1-d(t)sin(\theta_mt)\right]\\
        +j\cdot
        &\sqrt{T(t)\eta_e}A_p^m(t)\left[1+(d(t)cos(\theta_mt)-1)\right].
    \end{split}
\end{equation}

After matching filtering, that is, the same RRC filter as the transmitter, is applied to $\mu^{\rm{CD}}_3$ to get the recovered pilot signal $\mu^{\rm{MF}}_3$.
\begin{equation}
    \label{eq2.2-20}
    \begin{split}
        \mu^{\rm{MF}}_3 
       =&\sqrt{T_i\eta_e}A_p(1+j)+\varepsilon_p,
    \end{split}
\end{equation}
in which, 
\begin{equation}
    \label{eq2.2-21}
    \varepsilon_p=\sqrt{T_i\eta_e}A_p[-d_isin(\theta_m^i)+j(d_icos(\theta_m^i)-1)].
\end{equation}

$\varepsilon_p$ is defined as noise associated with modulation imbalance in the pilot. We can take out the part $\mu_4$ without noise.
\begin{equation}
    \label{eq2.2-22}
    \mu_4=\sqrt{T_i\eta_e}A_p(1+j).
\end{equation}

\begin{figure*}[!ht]
\centering
\subfigure[]
{
\includegraphics[width=0.95
\textwidth]{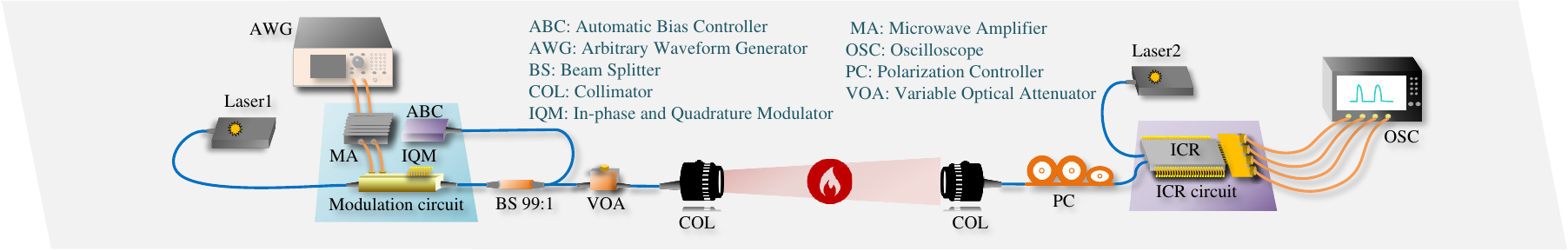}
}
\subfigure[]
{
\includegraphics[width=0.95\textwidth]{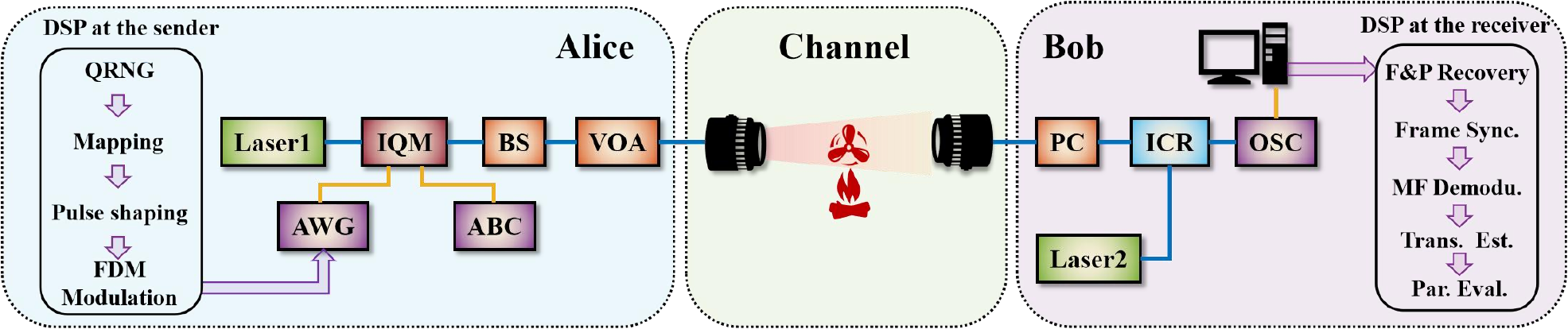}
}
\caption{(a) Optical layout of free-space LLO-CVQKD system. The red flame indicates that candles have been placed at the bottom of the free-space channel and are blown by a fan to simulate the effects of outdoor turbulence on channel transmittance. The blue line represents the optical signal channel, while the yellow line denotes the electrical signal channel. (b) Schematic diagram of the free space LLO-CVQKD experiment. In the sender's DSP, QRNG indicates the generation of random numbers. Mapping indicates mapping random numbers to random complex number sequences with Gaussian distribution. Pulse shaping means performing pulse shaping on random complex number sequences. FDM Modulation means combining the pilot and quantum signals in a frequency division multiplexing manner. In the DSP operations at the receiver, first, the time-varying recovery of frequency offset and phase (F\&P Recovert) for the received signal is carried out. After frame synchronization (Frame Sync.) and matched-filter demodulation (MF Demodu.) are completed, the time-varying transmittance is estimated (Trans. Est.). Finally, the relevant parameters are evaluated (Par. Eval.).}
\label{fig:3.1-1}
\end{figure*}

In the pre-calibration process, the pilot processes the same DSP operation as above, and we can get $\mu^0_4$ is
\begin{equation}
    \label{eq2.2-23}
    \mu^0_4=\sqrt{\eta_e}A_p(1+j).
\end{equation}

To get the transmittance $T_i$, all we need is
\begin{equation}
    \label{eq2.2-24}
    T_i=\frac{|\mu_4|^2}{|\mu^0_4|^2}=\frac{2T_i\eta_eA^2_p}{2\eta_eA^2_p}.
\end{equation}

Since the pilot and the quantum signal experience the same physical process, this is equivalent to the situation where the main quantum system and the auxiliary quantum system are both subjected to the same physical processes. Thus the estimated time-varying transmittance $T_i$ is applied equally to both. 

After DSP, the quantum measurement of the pilot achieves ideal mode matching. Furthermore, under the condition that ensures the signal-to-noise ratio of the pilot $\rm{SNR_p}$ is sufficiently high, the quantum signal can be successfully recovered using the DSP of the auxiliary quantum system. Therefore, the optimal DSP algorithm is 
\begin{equation}
    \label{eq2.2-25}
    \begin{split}
        f^{N,\rm{opt}}_{\rm{dsp}}=&\frac{\xi^0_{s}(t)}{\xi^0_{\rm{LO}}(t)}e^{j[\omega_{\rm{LO}}(t)t-\omega_s(t)t-\theta_{\rm{LO}}(t)+\theta_s(t)]}\\
        \overset{\rm{CW}}{=}&e^{j[\omega_{\rm{LO}}(t)t-\omega_s(t)t-\theta_{\rm{LO}}(t)+\theta_s(t)]}\\
        =&\mu_2.
    \end{split}
\end{equation}

Here, $\overset{\rm{CW}}{=}$ means that under continuous-wave conditions, $\frac{\xi^0_{s}(t)}{\xi^0_{\rm{LO}}(t)}$ is a fixed value and can be considered 1. Thus, the optimal DSP can be regarded as $\mu_2$.

Using the optimal DSP algorithm, the mode-matching coefficient of the quantum signal is
\begin{equation}
    \label{eq2.2-26}
    \begin{split}
        \sqrt{\eta_q}=&\int_{t_0}^{t_0+N/F_m}dt\frac{1}{\sigma_{\rm{cal}}^*}\xi_{\rm{LO}}^*(t)\xi_s(t)G_{\rm{dsp}}^{N,\rm{opt},*}(t)\\
        =&\int dt\int_{t_0}^{t_0+N/F_m}\frac{1}{\sigma_{\rm{cal}}^*}\xi_{\rm{LO}}^*(t)\xi_s(t)\frac{\xi_s^{0^*}(\tau)}{\xi_{\rm{LO}}^{0^*}(\tau)}\\
        &exp\left\{j\left[\omega_s(\tau)\tau-\omega_{\rm{LO}}(\tau)\tau+\theta_{\rm{LO}}(\tau)-\theta_s(\tau)\right]\right\}\\
        =&\int dt \int_{t_0}^{t_0+N/F_m} \frac{1}{\sigma_{\rm{cal}}^*}\xi_s^0(t)\xi_s^{0^*}(\tau)\\
        =&1.
    \end{split}
\end{equation}

\begin{figure*}[!ht]
\centering
\subfigure[]
{
\includegraphics[width=0.32
\textwidth]{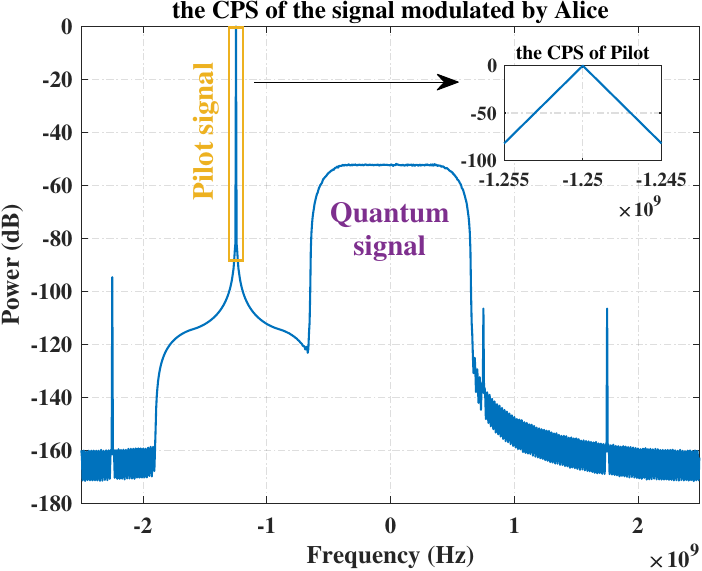}
}
\subfigure[]
{
\includegraphics[width=0.32\textwidth]{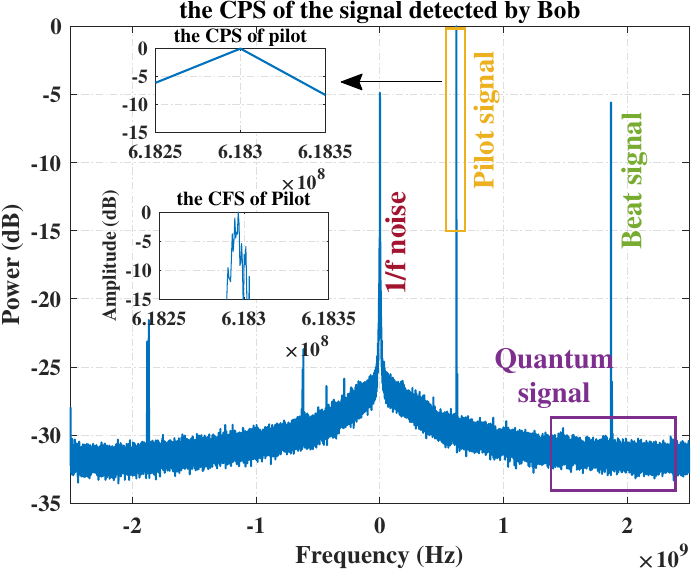}
}
\subfigure[]
{
\includegraphics[width=0.32\textwidth]{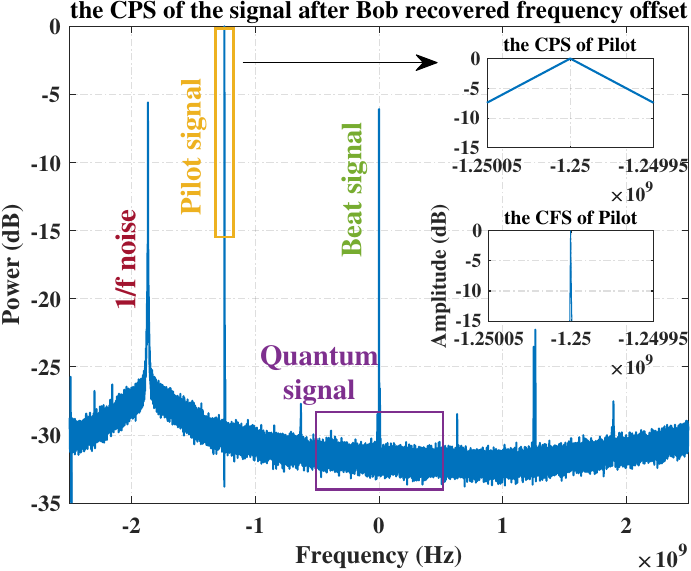}
}
\caption{Pictures related to DSP processing in free space LLO-CVQKD experiment. (a) The complex power spectrum (CPS) of the signal modulated by Alice. The subgraph is the CPS of the modulated pilot signal. (b) The CPS of the signal detected by Bob. The subgraphs are, respectively, the CPS and complex frequency spectrum (CFS) of the pilot detected by Bob. (c) The CPS of the signal after Bob recovered frequency offset. The subgraphs are, respectively, the CPS and CFS of the pilot signal after frequency offset recovery.}
\label{fig:3.2-1}
\end{figure*}

Furthermore, the first-order moment of the quantum signal is defined as follows:
\begin{equation}
    \label{eq2.2-27}
    \begin{split}
        \mu_q =& \left\langle\gamma_{\rm{quan}}\right|\sqrt{\eta_e\eta_q}\left(\hat{X}_{\xi_{\rm{s1}}}+j\cdot\hat{P}_{\xi_{\rm{s1}}}\right)\left|\gamma_{\rm{quan}}\right\rangle_{\xi_{\rm{s1}}}\\
        =&\sqrt{T(t)\eta_e}\left[x_q^m(t)+j\cdot p_q^m(t)\right]+\varepsilon_q\\
        \overset{\rm{MF}}{=}&\sqrt{T_i\eta_e}\left[x_q(i)+j\cdot p_q(i)\right]+\varepsilon_{\rm{q1}},
    \end{split}
\end{equation}
in which, the noise related to modulation imbalance in the quantum signal is
\begin{equation}
    \label{eq2.2-28}
    \varepsilon_{\rm{q1}}=\sqrt{T_i\eta_e}p_q(i)\left[-d_isin(\theta_m^i)+j(d_icos(\theta_m^i)-1)\right],
\end{equation}
and the second-order moment of the quantum signal is
\begin{equation}
    \label{eq2.2-29}
    \begin{split}
        \sigma^2_q=&\left\langle\gamma_{\rm{quan}}\right|\hat{I}_{\xi_{\rm{dsp}}}^{\rm{SNU}}\hat{I}_{\xi_{\rm{dsp}}}^{\rm{SNU},\dagger}\left|\gamma_{\rm{quan}}\right\rangle_{\xi_{\rm{s1}}}\\
        =&T_i\eta_eV_A+\rm{Var}(\varepsilon_{\rm{q1}})+1.
    \end{split}
\end{equation}

Here, $\rm{Var}()$ is the variance operation. In this way, we have accomplished the recovery of the quantum signal under the time-varying model. The effect introduced by modulation imbalance can be equivalent to 
noise. We also analyzed the impact of this noise in Appendix~\ref{Appendix:A5}. More importantly, the time-varying channel transmittance is also estimated.

\section{\label{section:3} Experimental setup}

\subsection{\label{section:3.1} Optical layout}

The experimental setup is illustrated in Fig.~\ref{fig:3.1-1}(a). The signal light is generated by Laser 1, a narrow linewidth laser that emits light at a central wavelength of 1550.12 nm. A 16-bit arbitrary waveform generator (AWG) outputs modulated electrical signals at a sampling rate of 5 Gs/s, which controls the IQM for continuous waveform modulation of the signal light from Laser 1. The pilot signal is obtained by filtering the DC signal with an RRC filter and then loading it on the carrier. The quantum signal is a pulse-shaped Gaussian-modulated coherent state (GMCS), also shaped by the same RRC filter. The amplitude ratio \( A_p \) between the pilot and quantum signals is 16. Both signals are transmitted through the channel using FDM, with the pilot signal modulated at a carrier frequency of \( f_p = -1.25 \) GHz, while the quantum signal is modulated at the baseband.

Key parameters of the modulation scheme include a modulation variance \( V_A \) of 13.78, a modulation frequency \( F_m \) of 1 GHz, and a roll-off factor of 0.3 for the RRC filter. After modulation, the light signal from the IQM passes through a 90:10 BS. Ten percent of the light is used to adjust the IQM bias point via an ABC system, while the remaining ninety percent is sent through a VOA. Following appropriate attenuation, the signal is coupled into the transmitter collimator and transmitted to Bob via free space. Several candles are positioned in the free-space channel between Alice and Bob to simulate atmospheric turbulence, and a fan is employed to induce movement in the flames \cite{ref20}. Laser 2 generates the LO light with a central wavelength of 1550.1350 nm. 

Bob's collimator facilitates the coupling of the received signal into the photodetector. After polarization adjustment, the signal undergoes heterodyne detection using the ICR. The AWG and the oscilloscope (OSC) achieve clock synchronization, and DSP is performed on the signal sampled by the OSC. Perform continuous measurements on each signal frame, with each frame containing $10^6$ ADC symbol samples. Utilizing back-to-back measurements, Alice and Bob are connected using a short optical fiber, the VOA is adjusted to achieve an appropriate $V_A$, and the signal is captured at this point as the pre-calibrated signal for estimating the time-varying transmittance of each point in a frame. 

Fig.~\ref{fig:3.1-1}(b) depicts the DSP flow for both the sender and receiver in the LLO-CVQKD system. At the sender's side, a quantum random number generator (QRNG) produces random numbers, which are then mapped to generate a sequence of complex numbers with a Gaussian distribution. This sequence represents the amplitude of the quantum signal used for IQM. The quantum signal operates at a symbol rate of 1 Gs/s, which is subsequently upsampled to 5 Gs/s. The upsampled signal undergoes pulse shaping using RRC filter with a roll-off factor of 0.3. After pulse shaping by the same RRC filter, the pilot signal is frequency-division multiplexed at -1.25 GHz. Bob uses an OSC to collect the detection signal and then performs DSP post-processing operations. This pilot signal plays a crucial role in the receiver's DSP operations, facilitating the time-varying recovery of frequency offset and phase noise, and the estimation of time-varying channel transmittance.

\subsection{\label{section:3.2} Time-varying high precision data processing}

\begin{figure*}[ht!]
    \centering
    \includegraphics[width=1\textwidth]{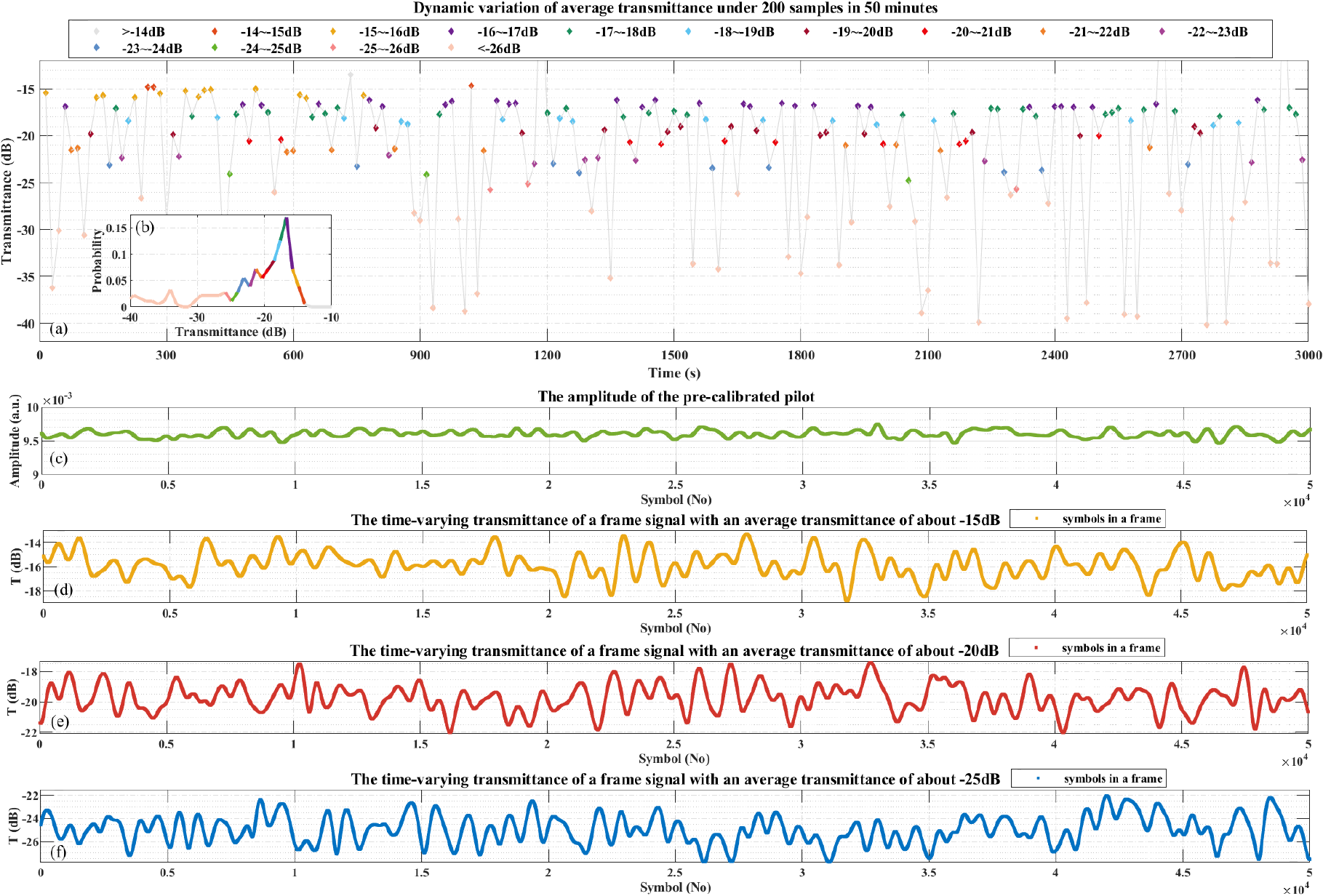}
    \caption{Transmittance of free space channel in LLO-CVQKD system. (a) The average transmittance of 200 frames was sampled in 50 minutes. Signs of different colors represent the transmittance of different intervals. (b) Probability distribution of channel transmittance. (c) The complex amplitude of the pilot signal during pre-calibration. The change of the complex amplitude is on the order of $10^{-3}$. We can consider the pilot amplitude to remain unchanged during pre-calibration. (d)-(f) Corresponding to the real-time transmittance in a frame when the average transmittance is at about -15, -20, and -25 dB, respectively.
}
    \label{fig:3.2-2}
\end{figure*}

The complex power spectrum (CPS) of the signal transmitted by Alice is illustrated in Fig.~\ref{fig:3.2-1}(a). The center frequency of the pilot signal is at -1.25 GHz, while the quantum signal is baseband modulated, centered at 0 GHz. Fig.~\ref{fig:3.2-1}(b) shows the CPS of the signal received by Bob, which is influenced by the frequency difference of the lasers between transmitter and receiver. Compared to the transmitted signal's CPS, the received signal experiences a frequency drift, causing the center frequency of the pilot signal to shift to approximately 0.6183 GHz. Additionally, a new signal spectrum appears on the negative half-axis. The positions of the latest and pilot signals are symmetrically distributed concerning the beat signal, resulting in modulation imbalance. Furthermore, 1/f noise is observed near the zero frequency, while the beat frequency generated by coherent detection is located at the center of the quantum signal spectrum. 

The inset in Fig.~\ref{fig:3.2-1}(b) corresponds to the CPS and complex frequency spectrum (CFS) of the pilot tone, demonstrating that the spectral shape of the pilot tone is no longer an impulse but has become irregularly broadened. However, this broadening phenomenon, which is caused by the time-varying drift of the laser central frequency and the time-varying system phase noise, is not evident in the CFS. It is important to note that spectral broadening also occurs in both the quantum and beat signals.

In the experiment, RRC shaping is carried out on the pilot signal to ensure that the $\rm{SNR_p}$ at the receiver could satisfy the time-varying recovery conditions. Therefore, through the Eqs.~(\ref{eq2.2-15}), (\ref{eq2.2-16}) and (\ref{eq2.2-25}) in theoretical analysis, the time-varying frequency offset and phase noise of quantum signal can be compensated. Fig.~\ref{fig:3.2-1}(c) shows the CPS of the signal after time-varying recovery. The center frequency of the pilot signal has been restored to -1.25 GHz, and the quantum signal has also returned to the baseband position. The pilot signal has also been converted to impulse response in the frequency domain. The beat frequency at 0 Hz within the signal spectrum is a DC component, which can be subsequently removed directly through appropriate operations.

\begin{figure*}[ht!]
    \centering
    \subfigure[]{
        \includegraphics[width=0.493\textwidth]{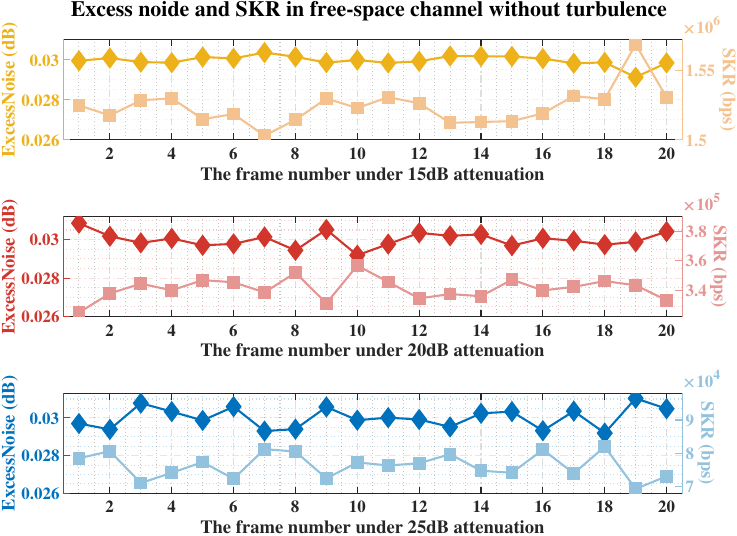}
        \label{label_for_cross_ref_1}
    }
    \subfigure[]{
	\includegraphics[width=0.46\textwidth]{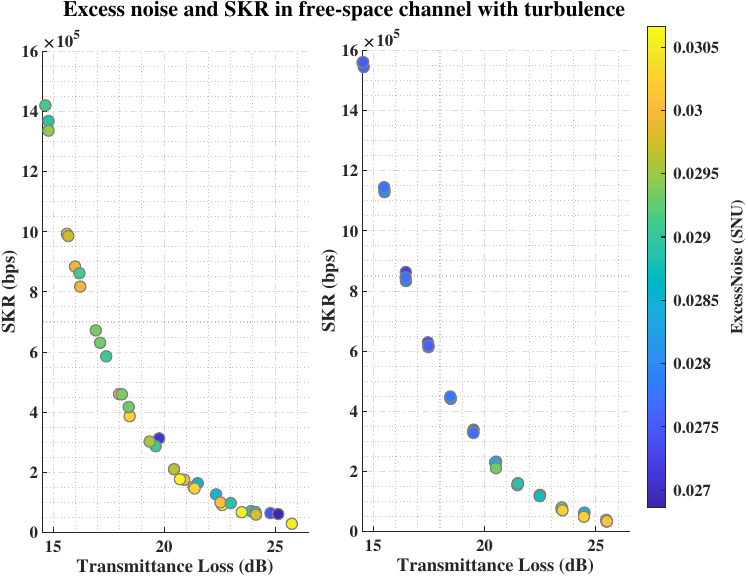}
        \label{label_for_cross_ref_2}
    }
    \caption{(a) In the channel without turbulence disturbance, when the channel transmittances are around 15 dB, 20 dB, and 25 dB, respectively, the excess noise and SKR of the system. (b) When a burning candle and a fan are added to the channel to simulate turbulence, the relationship among total system excess noise, channel transmission loss, and SKR.}
    \label{fig:4-1}
\end{figure*}

Next, the pilot and quantum signals are separated via band-pass and low-pass filters, and then matched-filter demodulation is performed to obtain the recovered signal. Consistent with the theoretical Eqs.~(\ref{eq2.2-27})-(\ref{eq2.2-29}), we equivalently treat the impact of modulation imbalance as noise. Appendix~\ref{Appendix:A5} analyzes the variation of the modulation imbalance noise (MIN) with the quadrature angle offset. Results show that within a deviation angle range of ±5°, the maximum value of MIN is 0.02 SNU. Fig.~\ref{fig:A6-1} in Appendix~\ref{Appendix:A6} shows the distribution histogram of the finally recovered quantum signal, which is well suited to the Gaussian distribution characteristics and explains the effectiveness of the recovery algorithm from another aspect.

More significantly, our study is the first to estimate the transmittance varying in time at each point within a signal frame. Specifically, we take 200 frame samples of the free-space channel simulating turbulence in 50 minutes, and thereby obtain the average transmittance for each frame sample, as depicted in Fig.~\ref{fig:3.2-2}(a). Due to the time interval between samples, the transmittance exhibits discontinuous jumps between -40 and -10 dB. Notably, in some instances, the detector's optical intensity is higher than that in non-turbulence measurements. This is due to the substantial common-mode noise from candlelight, rendering the noise of this signal too high for use.   Statistical analysis of channel transmittance yields the representative probability distribution of the channel transmittance shown in Fig.~\ref{fig:3.2-2}(b), which satisfies the negative logarithmic Weibull distribution. Appendix~\ref{Appendix:A7} theoretically analyzes that this distribution is close to the theoretical effect of turbulence on transmittance in the 10.5 km atmospheric channel \cite{ref49,ref50}.

Since Alice and Bob are connected by a very short optical fiber in the back-to-back calibration process, it can be considered that there is no channel attenuation in this case; that is, the channel transmittance is 0 dB. Fig.~\ref{fig:3.2-2}(c) shows the recovered pilot complex amplitude during back-to-back calibration. The complex amplitude only changes on the order of $10^{-3}$. At this time, the power of each pilot symbol can be considered unchanged and the pilot can be used as a pre-calibrated pilot in time-varying transmittance. Under the influence of the channel transmittance, the pilot signal's power obtained through coherent demodulation in the experimental environment will be reduced compared with the pilot signal recovered during the pre-calibration. Under the guidance of Eq.~(\ref{eq2.2-24}), the time-varying transmittance can be estimated by calculating the power change of each pilot symbol. Fig.~\ref{fig:3.2-2} (d), Fig.~\ref{fig:3.2-2} (e) and Fig.~\ref{fig:3.2-2} (f), respectively, show the time-varying channel transmittance changes in a frame when the average transmittance is about -15, -20 and -25 dB. Points within a frame are sampled continuously and the estimated real-time transmittance is continuously changing without any abrupt changes. The beam wandering and channel fluctuations caused by atmospheric turbulence can lead to fading noise \cite{ref31}. Appendix~\ref{Appendix:A8} gives the statistical characteristics of this noise. The results showed that the probability of fading noise is less than 0.003SNU. This means that there is less security risk caused by channel jitter.

\section{\label{section:4} Experimental results}

According to the measured excess noise, the asymptotic SKR of this setup can be calculated as follows. The excess noise is measured on a block of $1\times10^6$.
\begin{equation}
    \label{eq4-1}
    R=F_m(1-\rm{FER})\left[\beta I_{\rm{AB}}-\kappa_{\rm{BE}}-\Delta(n)\right],
\end{equation}

\noindent where $F_m$ is the repetition frequency of the CVQKD system, FER is the frame error rate of reverse reconciliation, $\beta$ is the reconciliation efficiency, $I_{\rm{AB}}$ is the Shannon mutual information between Alice and Bob, $\kappa_{\rm{BE}}$ is the Holevo bound of the information between Bob and Eve, and $\Delta(n)$ is related to the security of privacy amplification. The specific calculation process is in Appendix~\ref{Appendix:A9}. In the context of infinite size, $\Delta(n)$ tends to be 0.

When no channel turbulence disturbance is added, the channel attenuation is 15 dB, 20 dB, and 25 dB, respectively, as shown in Fig.~\ref{fig:4-1}(a). The scattered points corresponding to the left axis in the figure represent the magnitude of the excess noise at Alice. The value of the excess noise is the result of a reverse calculation from Bob's data. The repetition frequency of the modulated quantum signal is 1 GHz, the detection efficiency of the detector is 0.56, and the modulation variance is 12.4 SNU. When the channel attenuation is 15 dB, 20 dB, and 25 dB, the experimentally measured average excess noise values are approximately 0.029 SNU, 0.0300 SNU, and 0.029 SNU, respectively. The system repetition frequency is 1 GHz, FER with $\beta=0.96$ is obtained to 30\%, and the electrical noise $\nu_{\rm{el}}$ is 0.1 SNU. The scattered points corresponding to the right axis in Fig.~\ref{fig:4-1}(a) represent the magnitude of the SKR under this parameter setting. The calculated average key rates are 1.524 Mbps, 341.058 kbps, and 76.366 kbps, respectively. 

We constructed an experimental scenario to simulate CVQKD in free space affected by atmospheric turbulence. Specifically, we placed a lit candle in the communication channel. We used a fan to blow the candle flame, mimicking the effects of actual atmospheric turbulence on the free-space channel. The $V_A$ was set to 13.78 in our experimental setup, and the channel transmittance was approximately -14 dB without turbulence. Our experimental results indicate that the proposed scheme has a maximum channel attenuation tolerance of approximately -26 dB to evaluate the channel's critical rate effectively.

The final key rate in the asymptotic case is calculated according to the measured transmittance and the probability of the transmittance distribution, which can be expressed as
\begin{equation}
    \label{eq4-2}
    {R}_{\rm{final}}=\sum_{T=-26}^{T=-14} P_TR_{T},
\end{equation}
in which $P_T$ is the probability of the transmittance distribution, $R_{T}$ is obtained by calculating from Eq.~(\ref{eq4-1}), and the parameters $F_m$, FER, $\beta$ and $\nu_{\rm{el}}$ are the same as those used above. We calculate the key rate $R_{T}$ using two methods: one based on the average transmittance and the other based on the time-varying transmittance.

In the first scenario, we select signals with an average transmittance between -26 and -14 dB. Then, this average transmittance is used to determine the key rate for each frame signal, where $P_T$ represents the probability that the average transmittance occurs in each sub-interval. The left figure in Fig.~\ref{fig:4-1}(b) illustrates the relationship between transmittance loss, key rate, and excess noise derived from this calculation. The key rate decreases with the increase of channel attenuation. The average SKR calculated in this way is 363.498 kbps.

In the second scenario, we divide the range between -26 dB and -14 dB into 12 sub-intervals, each 1 dB wide. We subsequently ascertain the specific sub-interval for each signal by referencing the real-time transmittance associated with each data point, thereby generating 12 distinct sets of new signals. We then extract new frames with a length of $1\times10^6$ and calculate the key rate. $P_T$ represents the probability that the time-varying transmittance occurs within each sub-interval. The right figure in Fig.~\ref{fig:4-1}(b) illustrates the data relationship using this method. Compared to the first case, the excess noise calculated from real-time transmittance is slightly lower than the average transmittance. At the same time, the key rate is marginally higher than the average transmittance. The average SKR calculated in this way is 403.896 kbps.

\section{\label{section:5} Conclusion}

In this paper, we present a demonstration of simulated turbulent free-space CVQKD based on the LLO scheme. Considering time-varying parameter estimation, our experimental system achieves lower excess noise under high-frequency and high-attenuation conditions. We first discuss the influence of time-varying parameters on signal recovery in the CVQKD system. Subsequently, based on the theoretical principles of the time mode of continuous modes, we derive the theoretical basis of the time-varying parameter estimation scheme. We verified this scheme in a laboratory environment. We obtained the security key rate at transmittance from -14 dB to -26 dB in 1 dB units, and finally calculated the time-varying transmittance to get an average key rate of 403.896 kbps. This provides a foundation for key distribution between satellites and the ground under greater attenuation.

The above theoretical and experimental evidence shows we can achieve free-space quantum key distribution under large attenuation. However, for simplicity in the derivation, theoretically, we only modeled and analyzed the detector's detection efficiency and electrical noise, without considering other effects like nonlinear effect. In the experiment, these effects are all treated as noise. For practical field-test experiments, polarization changes will also impact excess noise, which is not considered in our research. Additionally, we have not temporarily considered the influence of the finite block size effect. In future work, we will consider more sophisticated models and more complex scenarios to achieve transmission under greater attenuation with larger block sizes in practical field-test experiments.

\begin{acknowledgments}
This work was supported by the Innovation Program for Quantum Science and Technology (Grant No. 2021ZD0300703), Shanghai Municipal Science and Technology Major Project (2019SHZDZX01), the Key R\&D Program of Guangdong province (Grant No. 2020B0303040002), and the National Natural Science Foundation of China (No. 62101320).
\end{acknowledgments}

\section*{Data Availability Statement}

The data that support the findings of this study are available from the corresponding author upon reasonable request.

\appendix

\section{\label{Appendix:A1} Fundamental definitions for temporal modes of continuous-mode states}

We can obtain the basic definition of the temporal modes in a continuous mode according to Refs.~\onlinecite{ref42,ref43,ref44,ref45,ref46,ref47}. The creation operator $\hat{a}^{\dagger}(\omega)$ and annihilation operator $\hat{a}(\omega)$ in continuous mode can be obtained by transforming the corresponding operators $\hat{a}^{\dagger}_{i}$ and $\hat{a}_{i}$ in discrete mode. In continuous mode, the commutation relation is still satisfied, that is, $\left[\hat{a}(\omega_{i}),\hat{a}^{\dagger}(\omega_{j})\right]=\delta_{ij}$, where $\delta()$ is Dirac function. In the time domain, the creation operator $\hat{a}^{\dagger}(t)$ and $\hat{a}^{\dagger}(\omega)$ are Fourier transforms of each other, and the relationship of the annihilation operators is the same.

The initial envelope in the i-th period is defined as $\xi^{0}_{i}(t)$. The wavepacket $\xi_{i}(t)$ carrying an optical carrier with frequency $\omega(t)$ is defined as $\xi^{0}_{i}(t)e^{-j\omega(t)t}$. The wavepackets $\xi_{i}(t)$ and $\xi_{j}(t)$ in different time periods satisfy the orthogonal condition $\int dt\xi_{i}(t)\xi^{*}_{j}(t)=\delta_{ij}$. At this time, the photon-wavepacket operators can be jointly defined by the wavepacket and the continuous mode operator as follows:
\begin{equation}
    \label{eqA1-1}
    \begin{split}
        &\hat{A}_{\xi_{i}}=\int dt\xi^*_{i}(t)\hat{a}(t),\\
        &\hat{A}^{\dagger}_{\xi_{i}}=\int dt\xi_{i}(t)\hat{a}^{\dagger}(t).
    \end{split}
\end{equation}

In this context, the orthogonal operators can be defined as 
\begin{equation}
    \label{eqA1-2}
    \begin{split}
&\hat{X}_{\xi_{i}}=\hat{A}^{\dagger}_{\xi_{i}}+\hat{A}_{\xi_{i}},\\
&\hat{P}_{\xi_{i}}=j\left(\hat{A}^{\dagger}_{\xi_{i}}-\hat{A}_{\xi_{i}}\right).
    \end{split}
\end{equation}

\section{\label{Appendix:A2} Basic operators in temporal modes of continuous-mode states}
The displacement operator in continuous mode generated by the linear Hamiltonian is defined as \cite{ref43}
\begin{equation}
    \label{eqA2-1}
    \hat{D}_{\xi_{i}}(\gamma)=exp\left\{\gamma\hat{A}^{\dagger}_{\xi_{i}}-\gamma^{*}\hat{A}_{\xi_i}\right\}.
\end{equation}

Here, $\gamma$ is a complex number representing the displacement parameter acting on the photon-wavepacket. In the Heisenberg picture, the continuous-mode annihilation operator after the linear unitary Bogoliubov transformation is $\hat{A}_{\xi_i}\xrightarrow{}\hat{A}_{\xi_i}+\gamma$ \cite{ref48}. Applying the displacement operator $\hat{D}_{\xi_{i}}(\gamma)$ to the vacuum state $\left|0\right\rangle$ can generate the photon-wavepacket coherent state $\left|\gamma\right\rangle_{\xi_{i}}$. 
\begin{equation}
    \label{eqA2-2}
    \begin{split}
        \left|\gamma\right\rangle_{\xi_{i}}=\hat{D}_{\xi_{i}}(\gamma) \left|0\right\rangle 
        =exp\left\{\gamma\hat{A}^{\dagger}_{\xi_{i}}-\gamma^{*}\hat{A}_{\xi_{i}}\right\}\left|0\right\rangle.
    \end{split}
\end{equation}
$|\gamma|^2$ is the average photon number. The photon-wavepacket coherent state satisfies the eigenvalue equation $\hat{A}_{\xi_{i}}\left|\gamma\right\rangle_{\xi_{i}}=\gamma\left|\gamma\right\rangle_{\xi_{i}}$. 

The phase rotation operator in continuous mode defined by the free propagation Hamiltonian is $\hat{R}_{\xi_i}(\theta)=exp\left\{-j\theta\hat{A}^{\dagger}_{\xi_i}\hat{A}_{\xi_i}\right\}$ \cite{ref48}, where $\theta$ is the phase rotation of the photon-wavepacket. In the Heisenberg picture, after the annihilation operator undergoes a linear unitary Bogoliubov transformation, $\hat{A}_{\xi_i}\xrightarrow{}e^{-j\theta}\hat{A}_{\xi_i}$ \cite{ref48}. Apply the phase rotation operator to the coherent state $\left|\gamma\right\rangle_{\xi_{i}}$
\begin{equation}
    \label{eqA2-3}
    \begin{split}
        \hat{R}_{\xi_i}(\theta)\left|\gamma\right\rangle_{\xi_i}
        =exp\left\{\gamma e^{j\theta}\hat{A}^{\dagger}_{\xi_{i}}-\gamma^{*}e^{-j\theta}\hat{A}_{\xi_{i}}\right\}\left|0\right\rangle
        =\left|\gamma e^{j\theta}\right\rangle_{\xi_i}.
    \end{split}
\end{equation}
 
The transformation of the beam splitter in continuous mode is defined as \cite{ref48}
 \begin{equation}
     \label{eqA2-4}
     \hat{B}_{\xi_i}(\tau)=exp\left\{\tau\left(\hat{A}^{\dagger}_{\xi_i}\hat{B}_{\xi_i}-\hat{A}_{\xi_i}\hat{B}^{\dagger}_{\xi_i}\right)\right\},
 \end{equation}
 in which, $\hat{A}_{\xi_i}$ and $\hat{B}_{\xi_i}$ are the annihilation operators of two modes, and $\tau$ is a parameter related to the transmittance $T$ of the BS, and their relationship satisfies $T=cos^{2}\tau$. In the Heisenberg picture, the annihilation operator is transformed into Eq.~(\ref{eqA2-5}) after undergoing a linear unitary Bogoliubov transformation \cite{ref48}. 
\begin{equation}
    \label{eqA2-5}
    \begin{bmatrix}
        \hat{A}_{\xi_i}\\
        \hat{B}_{\xi_i}
    \end{bmatrix}
    \xrightarrow{}
    \begin{bmatrix}
        \sqrt{T} & \sqrt{1-T}\\
        -\sqrt{1-T} & \sqrt{T}
    \end{bmatrix}
    \begin{bmatrix}
         \hat{A}_{\xi_i}\\
        \hat{B}_{\xi_i}
    \end{bmatrix}.
\end{equation}

Let the model act on the two coherent states $\left|\alpha\right\rangle_{\xi_{i}}$ and $\left|\beta\right\rangle_{\xi_{i}}$ which are input into the BS, and define  $R=1-T$. According to Appendix~\ref{Appendix:A1}, we can obtain
\begin{equation}
    \label{eqA2-6}
    \begin{split}
        &\hat{B}_{\xi_i}(\tau)\left|\alpha\right\rangle_{\xi_i}\otimes \left|\beta\right\rangle_{\xi_i}\\
        =&exp
        \begin{Bmatrix}
            \alpha\left(\sqrt{T}\hat{A}^{\dagger}_{\xi_i}+\sqrt{R}\hat{B}^{\dagger}_{\xi_i}\right)-
            \alpha^{*}\left(\sqrt{T}\hat{A}_{\xi_i}+\sqrt{R}\hat{B}_{\xi_i}\right)
        \end{Bmatrix}
        \left|0\right\rangle\otimes\\
        & exp
        \begin{Bmatrix}
            \beta\left(-\sqrt{R}\hat{A}^{\dagger}_{\xi_i}+\sqrt{T}\hat{B}^{\dagger}_{\xi_i}\right)-
            \beta^{*}\left(-\sqrt{R}\hat{A}_{\xi_i}+\sqrt{T}\hat{B}_{\xi_i}\right)
        \end{Bmatrix}
        \left|0\right\rangle\\
        =&exp
        \begin{Bmatrix}
            \left(\sqrt{T}\alpha-\sqrt{R}\beta\right)\hat{A}^\dagger_{\xi_i}-\left(\sqrt{T}\alpha^*-\sqrt{R}\beta^*\right)\hat{A}_{\xi_i}
        \end{Bmatrix}
       \left|0\right\rangle \otimes\\
       &exp
        \begin{Bmatrix}
           \left(\sqrt{R}\alpha+\sqrt{T}\beta\right)\hat{B}^\dagger_{\xi_i}-\left(\sqrt{R}\alpha^*+\sqrt{T}\beta^*\right)\hat{B}_{\xi_i}
        \end{Bmatrix}
       \left|0\right\rangle\\
       =&\left|\sqrt{T}\alpha-\sqrt{R}\beta\right\rangle_{\xi_i}\otimes \left|\sqrt{R}\alpha+\sqrt{T}\beta\right\rangle_{\xi_i}.
    \end{split}
\end{equation}

After passing through the BS, two coherent states are divided into another two states: $\left|\sqrt{T}\alpha-\sqrt{R}\beta\right\rangle_{\xi_i}$ and $\left|\sqrt{R}\alpha+\sqrt{T}\beta\right\rangle_{\xi_i}$.

\section{\label{Appendix:A3} Mode-matching coefficient}
According to the photoelectric flux operator of the homodyne detection receiver \cite{ref46,ref47}, we can generalize to formulate the two orthogonal photoelectric flux operators pertinent to a heterodyne detection system as follows:

\begin{equation}
    \label{eqA3-1}
    \begin{split}
        \hat{f}_x(t)&=\left[\hat{a}_s^\dagger(t)\hat{a}_{\rm{LO}}(t)+\hat{a}_{\rm{LO}}^\dagger(t)\hat{a}_s(t)\right]\otimes g_x(t),\\
        \hat{f}_p(t)&=\left[\hat{a}_{\rm{LO}}^\dagger(t)\hat{a}_s(t)-\hat{a}_s^\dagger(t)\hat{a}_{\rm{LO}}(t)\right]\otimes g_p(t).
    \end{split}
\end{equation}

In this context, $g_x(t)$ and $g_p(t)$ represent two distinct filtering functions applied to the I and Q signals, respectively. $\hat{a}_s(t)$ and $\hat{a}_{\rm{LO}}(t)$ are the photon wavepackets of signal light and LO light, respectively. The average photoelectric flux is
\begin{equation}
    \label{eqA3-2}
    \hat{f}_{\rm{LO}}(t)=\frac{\left\langle\alpha_{\rm{LO}}(t)\right|\hat{f}_x(t)\left|\alpha_{\rm{LO}}(t)\right\rangle+\left\langle\alpha_{\rm{LO}}(t)\right|\hat{f}_p(t)\left|\alpha_{\rm{LO}}(t)\right\rangle}{2}.
\end{equation}

The $G_{\rm{dsp}}^N(t)$ related to the DSP algorithm $f_{\rm{dsp}}$ and the weighted sum of the impulse response function is expressed as
\begin{equation}
    \label{eqA3-3}
    {G_{\rm{dsp}}^N(t)=\sum_{m=1}^{N}f_{\rm{dsp}}^m\int_{t_0+(m-1)/F_m}^{t_0+m/F_m}g_x(\tau-t)+j\cdot g_p(\tau-t)d\tau}.
\end{equation}

The photon-wavepacket expression of the temporal mode $\xi_{\rm{dsp}}(t)$ is
\begin{equation}
    \label{eqA3-4}
    \xi_{\rm{dsp}}(t)=\frac{1}{\sigma_{\rm{cal}}}\xi_{\rm{LO}}(t)G_{\rm{dsp}}^N(t).
\end{equation}

Here, $\sigma_{\rm{cal}}=\sqrt{\int dt\left|\xi_{\rm{LO}}^0(t)\right|^2\left[G_{\rm{dsp}}^N(t)\right]^2}$ is the rescaled factor when calibrating output data by SNU. Therefore, the mode-matching coefficient can be expressed as
\begin{equation}
    \label{eqA3-5}
    \begin{split}
        \sqrt{\eta}=&\int_{t_0}^{t_0+N/F_m}dt\xi_{\rm{dsp}}^*(t)\xi_s(t)\\
        =&\int_{t_0}^{t_0+N/F_m}dt\frac{1}{\sigma_{\rm{cal}}^*}\xi_{\rm{LO}}^*(t){G_{\rm{dsp}}^{N^*}}(t)\xi_s(t).
    \end{split}
\end{equation}

Certainly, we can obtain the mode-matching coefficients for the I and Q paths independently by extracting the real and imaginary components of $\sqrt{\eta}$, respectively.

\section{\label{Appendix:A4} Moments of measured data}
Before deriving the calculation, briefly introducing several pertinent and useful relationships is necessary. Firstly, the photon-wavepacket operators satisfy the following commutation relation:
\begin{equation}
    \label{eqA4-1}
    \left[\hat{A}_{\xi_i},\hat{A}^{\dagger}_{\xi_i}\right]=\hat{A}_{\xi_i}\hat{A}^{\dagger}_{\xi_i}-\hat{A}^{\dagger}_{\xi_i}\hat{A}_{\xi_i}=1.
\end{equation}

Secondly, the coherent state of photon-wavepacket fulfills the subsequent eigenvalue equation:
\begin{equation}
    \label{eqA4-2}
    \hat{A}_{\xi_i}\left|\gamma\right\rangle=\gamma\left|\gamma\right\rangle.
\end{equation}

By fully utilizing the relationships above, we can define the first and second-order moments and derive the desired results. 

The first-order moment (mean value) is given by
\begin{equation}
    \label{eqA4-3}
    \begin{split}
        \mu =& \left\langle \gamma\right|X_{\xi_i}+j\cdot P_{\xi_i}\left|\gamma\right\rangle_{\xi_i}\\
        =&\left\langle \gamma
        \right|\hat{A}_{\xi_i}\left|\gamma\right\rangle_{\xi_i}\\
        =&\left\langle\gamma\right|\gamma\left|\gamma
        \right\rangle_{\xi_i}\\
        =&\gamma\left\langle\gamma\right.\left|\gamma
        \right\rangle_{\xi_i}\\
        =&\gamma.      
    \end{split}
\end{equation}

The second-order moment is given by
\begin{equation}
    \label{eqA4-4}
    \begin{split}
        \sigma^2=&\left\langle\gamma\right|\left(X_{\xi_i}+j\cdot P_{\xi_i}\right)\left(X_{\xi_i}+j\cdot P_{\xi_i}\right)^*\left|\gamma\right\rangle_{\xi_i}\\
        =&\left\langle\gamma\right|\hat{A}_{\xi_i}\hat{A}^{\dagger}_{\xi_i}\left|\gamma\right\rangle_{\xi_i}\\
        =&\left\langle\gamma\right|\hat{A}^{\dagger}_{\xi_i}\hat{A}_{\xi_i}+1\left|\gamma\right\rangle_{\xi_i}\\
        =&\left[\hat{A}_{\xi_i}\left|\gamma\right\rangle_{\xi_i}\right]^\dagger\gamma\left|\gamma\right\rangle_{\xi_i}+\left\langle\gamma\right.\left|\gamma
        \right\rangle_{\xi_i}\\
        =&\gamma\gamma^*\left\langle\gamma\right.\left|\gamma
        \right\rangle_{\xi_i}+1\\
        =&\left|\gamma\right|^2+1.
    \end{split}
\end{equation}
where $\left|\gamma\right|$ is the modulus of $\gamma$.

\section{\label{Appendix:A5} Noise of modulation imbalance}

According to Eq.~(\ref{eq2.2-28}), the variance of the modulation imbalance noise (MIN) in quantum signal at Bob's end is
\begin{equation}
    \label{eqA5-1}
    {\rm{Var}[\varepsilon_{\rm{q1}}]}_B=T\eta_e\frac{V_A}{2}\left[d^2sin^2(\theta_m)+(dcos(\theta_m)-1)^2\right].
\end{equation}

Converted to the variance of the sender, we get
\begin{equation}
    \label{eqA5-2}
    \begin{split}
    {\rm{Var}[\varepsilon_{\rm{q1}}]}_A=&\frac{{\rm{Var}[\varepsilon_{\rm{q1}}]}_B}{T\eta_e}\\
    =&\frac{V_A}{2}\left[d^2sin^2(\theta_m)+(dcos(\theta_m)-1)^2\right].
    \end{split}
\end{equation}

\begin{figure}
\includegraphics[width=0.45\textwidth]{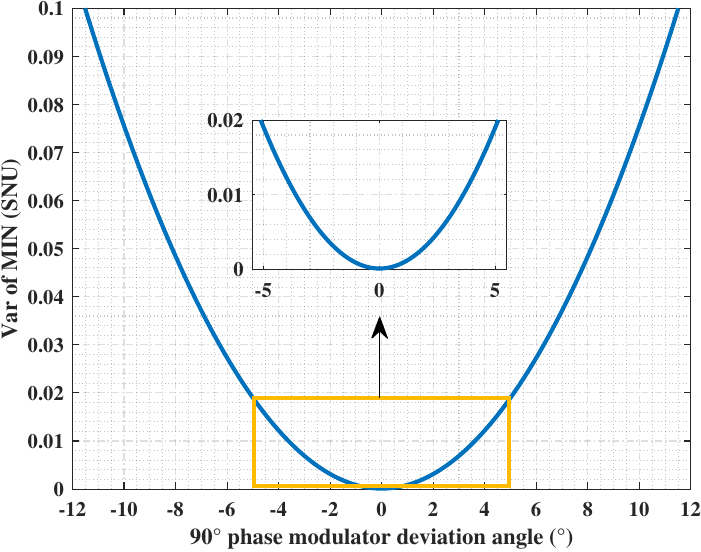}
\caption{The variance of modulation imbalance noise varies with PM deviation angle.}
\label{fig:A5-1}
\end{figure}

The minimum amplitude imbalance parameter $d_{\rm{min}}=1-10^{-22/10}=0.9937$, can be determined by analyzing the pilot signal and the symmetrical signal resulting from the imbalance, as shown in Fig.~\ref{fig:3.2-1}. Additionally, Fig.~\ref{fig:A5-1} illustrates how the offset angle of the 90° phase modulator in the IQ modulator affects the variance of modulator imbalance noise ${\rm{Var}[\varepsilon_{\rm{q1}}]}_A$.

In the experiment, the angular offset range of the PM obtained from IQ modulator testing is -5° to 5°, and ${\rm{Var}[\varepsilon_{\rm{q1}}]}_A$ is significantly less than 0.02 SNU, which is less than the excess noise in the experiment, satisfying the secret key generation conditions.

\section{\label{Appendix:A6} Histogram of recovered signal}
\begin{figure}
\includegraphics[width=0.45\textwidth]{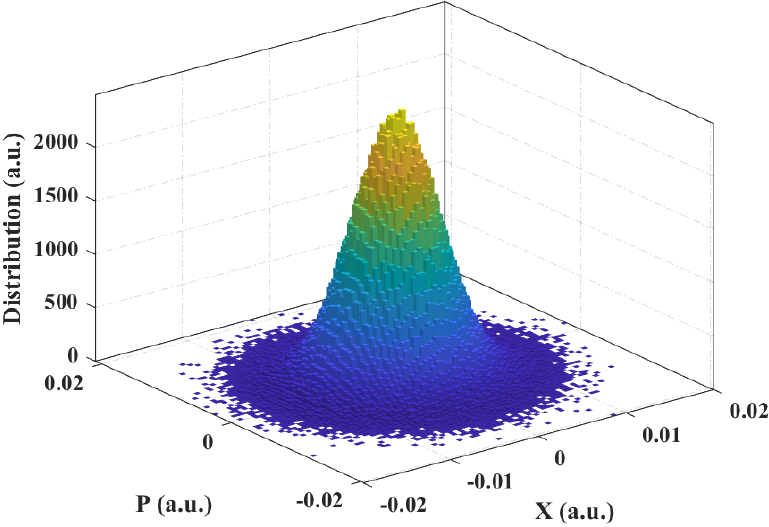}
\caption{The histogram of the recovered signal.}
\label{fig:A6-1}
\end{figure}

During our deep analysis of the signal recovered by the DSP, we created a statistical histogram (as shown in Fig.~\ref{fig:A6-1}). The shape of the histogram indicates that its distribution curve is approximately bell-shaped and unimodal, with the peak value at 0. This suggests that most of the data is concentrated around this point. The data symmetrically decreases on both sides of the peak, showing a balanced distribution.

Overall, the data exhibits clear characteristics of a Gaussian distribution, which further supports the effectiveness of the DSP recovery algorithm.

\section{\label{Appendix:A7} Simulation of equivalent turbulence}

\begin{figure*}
    \subfigure[]{
        \includegraphics[width=8cm]{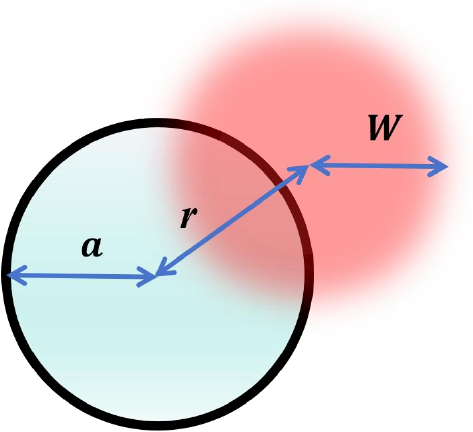}
        \label{label_for_cross_ref_3}
    }
    \subfigure[]{
	\includegraphics[width=8.5cm]{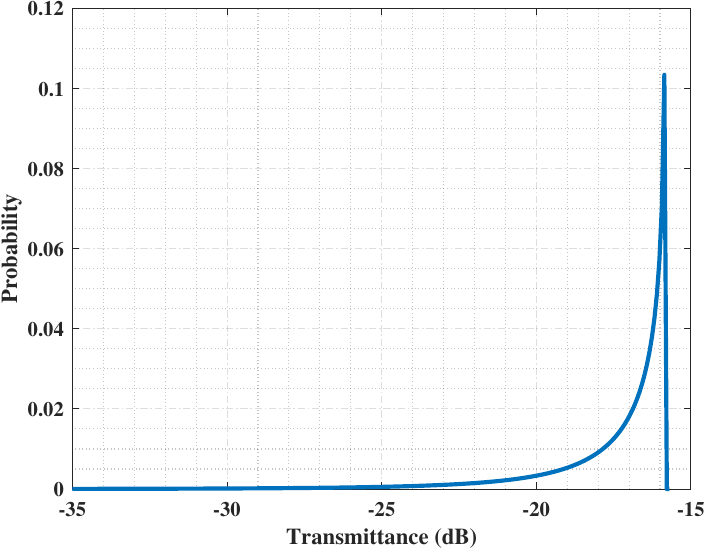}
        \label{label_for_cross_ref_4}
    }
    \caption{Free space channel characteristics. (a) A telescope with an aperture of $a$ captures light with a radius of $W$. (b) Probability distribution of transmittance of 10.5 km free space channel.}
    \label{fig:A7-1}
\end{figure*}
Atmospheric turbulence significantly impacts optical signal transmission, leading to random beam intensity, phase, and propagation direction fluctuations. These changes can degrade the performance of optical communication systems and undermine the security and reliability of quantum key distribution. Therefore, establishing an accurate equivalent turbulence simulation model is crucial.

Atmospheric turbulence arises from random fluctuations in atmospheric temperature, pressure, wind speed, and other factors, leading to unpredictable changes in the refractive index, distortions in light propagation paths, scattering, and energy loss. In a state of weak atmospheric turbulence, classical theory and experimental research indicate that the intensity of turbulence can be determined using a specific formula \cite{ref51}.
\begin{equation}
    \label{eqA7-1}
    \sigma^2\approx1.919C_n^2z^3(2W_0)^{-1/3}.
\end{equation}

In this formula, $C_n^2$ represents the structural parameter of the refractive index, which measures the intensity of atmospheric turbulence. Its value varies dynamically based on factors such as location, weather conditions, altitude, and time, necessitating estimation through field measurements or reliable meteorological data. $z$ refers to the propagation distance of the beam in free space, which depends on the design of the communication link and the specific application scenario. $W_0$ denotes the beam waist radius at the point where the beam enters the atmosphere, determined by the characteristics of the laser and the transmitting optical system.

The aperture of the receiving telescope is crucial for the efficiency and quality of the optical signal, and its interaction with the beam is complex. This interaction significantly affects the performance of signal acquisition and transmission. About the aperture model of the receiving telescope shown in Fig.~\ref{fig:A7-1}(a), let $r$ represent the distance between the center of the telescope's aperture and the center of the beam. The relationship between this distance $r$ and the transmittance $T^2$ is described as follows:
\begin{equation}
    \label{eqA7-2}
    T^2(r)=T_0^2exp\left[-\left(\frac rR\right)^\lambda\right].
\end{equation}

In this formula, $T_0^2$ represents the maximum transmittance determined by the beam radius \(W\) and the aperture radius \(a\). Its physical significance refers to the maximum ratio of the optical power received by the telescope to the total transmitted optical power, achievable under ideal alignment conditions when there are no other losses or interference. The parameters $\lambda$ and \(R\) are proportional and shape parameters, respectively, established through optical principles and mathematical derivation \cite{ref49}.
\begin{equation}
    \label{eqA7_3}
    T_0^2=1-exp\left[-2\frac{a^2}{W^2}\right],
\end{equation}
\begin{equation}
    \label{eq1208_4}
    \begin{split}
        \lambda=&8\frac{a^2}{W^2}\frac{exp[-4\frac{a^2}{W^2}]I_1(4\frac{a^2}{W^2})}{1-exp\left[-4\frac{a^2}{W^2}\right]I_0(4\frac{a^2}{W^2})}\\
        &\times\left[ln\left(\frac{2T_0^2}{1-exp[-4\frac{a^2}{W^2}]I_0(4\frac{a^2}{W^2})}\right)\right]^{-1},
    \end{split}
\end{equation}
\begin{equation}
    \label{eqA7-4}
    R=a\left[ln\left(\frac{2T_0^2}{1-exp[-4\frac{a^2}{W^2}]I_0(4\frac{a^2}{W^2})}\right)\right]^{-\frac1\lambda}.
\end{equation}

\noindent in which, $I_0()$ and $I_1()$ are modified Bessel functions. When the beam fluctuates around the aperture center, $T^2(r)$ satisfies the log-negative Weibull distribution and $P(T)$ is given by
\begin{equation}
    \label{eqA7-5}
    \begin{split}
        P(T)=\frac{2R^{2}}{\sigma^{2}\lambda T}\left(2ln\frac{T_{0}}{T}\right)^{\frac{2}{\lambda}-1}\times exp\left[-\frac{1}{2\sigma^{2}}R^{2}\left(2ln\frac{T_{0}}{T}\right)^{\frac{2}{\lambda}}\right],
    \end{split}
\end{equation}
for $T\in[0,T_0]$, and $P(T)=0$ else.

In the existing field environment experiment, the telescope aperture radius $a$ is 125 mm, and the beam waist $W_0$ before entering the atmospheric channel is 62.5 mm. The beam divergence angle is $5^{''}$, resulting in a beam waist of approximately 103.8 mm after passing through a 10.5 km free-space channel. Additionally, there is a 14 dB attenuation due to fiber coupling loss and lens absorption of light. We used the Eq.~(\ref{eqA7-5}) to obtain the probability distribution of  $T$, as shown in Fig.~\ref{fig:A7-1}(b). The simulated transmittance probability distribution aligns well with the experimental tolerance range, with both distributions being log-negative Weibull distributions. Therefore, using the existing equipment, conducting our experiment over at least a 10.5 km mild turbulent atmospheric channel is feasible. 

\section{\label{Appendix:A8} Fading noise}

In the real free space channel, the transmittance fluctuates due to atmospheric effects, and the channel attenuation is not fixed. Transmission fluctuations caused by beam wandering and atmospheric turbulence, resulting in fading noise, which can be calculated through \cite{ref31}
\begin{equation}
    \label{eqA8-1}
    \varepsilon_f={\rm{Var}}\left(\sqrt{T(t)}\right)\left(V_A-1\right),
\end{equation}
where the variance of the time-varying transmittance is
\begin{equation}
    \label{eqA8-2}
    {\rm{Var}}\left(\sqrt{T(t)}\right)=\overline{T(t)}-\overline{\sqrt{T(t)}}^2,
\end{equation}
where $\overline{T(t)}$ is the mean value of $T(t)$.

To calculate the probability distribution of fading noise, a continuous segment is randomly intercepted from the real-time transmittance continuously collected over a long period. The intercepted length $N_c$ is related to the length $N$ of one frame of signal, the acquisition time $t$ of one frame of signal, and the total acquisition time $t_{\rm{tot}}$, $N_c = N\cdot t/t_{\rm{tot}}$. The fading noise is calculated by using Eq.~(\ref{eqA8-1}). Repeating this process randomly several times can obtain the probability distribution of fading noise.

\begin{figure}
\includegraphics[width=0.45\textwidth]{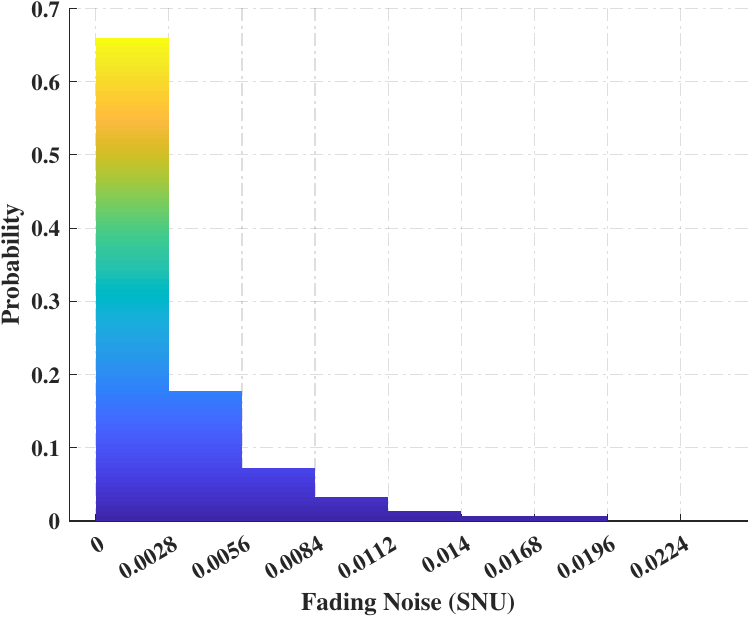}
\caption{Probability distribution of fading noise.}
\label{fig:A8-1}
\end{figure}

Fig.~\ref{fig:A8-1} illustrates the probability distribution of fading noise. The concentrated distribution of fading noise is below 0.003 SNU, indicating that the channel characteristics are more stable. This stability reduces the negative impact on mutual information, minimizes the risk of information leakage, and further enhances secure key distribution.

\section{\label{Appendix:A9} SKR calculation}
$I_{\rm{AB}}$ is calculated by the following formula
\begin{equation}
    \label{eqA9-1}
    \begin{split}
        I_{\rm{AB}}=&\log_2{\frac{V+\chi_{\rm{tot}}}{1+\chi_{\rm{tot}}}}\\
        =&\log_2{\frac{V+\chi_{\rm{line}}+\chi_{\rm{het}}/T}{1+\chi_{\rm{line}}+\chi_{\rm{het}}/T}}.
    \end{split}
\end{equation}

Here, $V=V_A+1$, $V_A$ is the modulation variance. $\chi_{\rm{tot}}$ is the total noise, $\chi_{\rm{line}}=1/T-1+\varepsilon$, and $\chi_{\rm{het}}=[1+(1-\eta)+2\mu_{\rm{el}}]/\eta$. Under asymptotic circumstances, T denotes the transmittance of the channel, and $\varepsilon$ represents Alice's excess noise.

$\kappa_{\rm{BE}}$ is described as
\begin{equation}
    \label{eqA9-2}
    \begin{split}
        \kappa_{\rm{BE}}=\sum_{i=1}^{2}G\left(\frac{\lambda_i-1}{2}\right)-\sum_{i=3}^{5}G\left(\frac{\lambda_i-1}{2}\right).
    \end{split}
\end{equation}

Here, $G(x)=(x+1)\log_2(x+1)-x\log_2x$. The value of $\lambda_i$ is a symplectic eigenvalue obtained from the covariance matrix and can be represented as
\begin{equation}
    \label{eqA9-3}
    \begin{split}
        &\lambda_{1,2}=\sqrt{\frac{1}{2}\left(A\pm{}\sqrt{A^2-4B}\right)},\\
        &\lambda_{3,4}=\sqrt{\frac{1}{2}\left(C\pm{}\sqrt{C^2-4D}\right)},\\
        &\lambda_5=1.
    \end{split}
\end{equation}

A, B, C, and D can be obtained through
\begin{equation}
    \label{eqA9-4}
    \begin{split}
        A&=V^2\left(1-2T\right)+2T+T^2\left(V+\chi_{\rm{line}}\right)^2,\\
        B&=T^2\left(V\cdot \chi_{\rm{line}}+1\right)^2,\\
        C&=\frac{1}{T^2\left(V+\chi_{\rm{tot}}^2\right)}\left(\left.A\cdot \chi_{\rm{het}}^2+B+1\right.\right.\\
        &\left.\left.+2\chi_{\rm{het}}\left(V\sqrt{B}+T(V+\chi_{\rm{line}})\right)+2T(V^2-1)\right)\right.,\\
        D&=\left(\frac{V+\sqrt{B}\chi_{\rm{het}}}{T(V+\chi_{\rm{tot}})}\right)^2.
    \end{split}
\end{equation}

Finally, $\Delta(n)$ is related to the security of the privacy amplification. It can be calculated as
\begin{equation}
    \label{eqA9-5}
    \begin{split}
        \Delta(n)=7\sqrt{\frac{\log_2(1/\overline{\varepsilon})}{n}}+\frac{2}{n}\log_2\frac{1}{\varepsilon_{\rm{pe}}},
    \end{split}
\end{equation}
in which, $\overline{\varepsilon}$ represents a smoothing parameter, and $\varepsilon_{\rm{pe}}$ denotes the failure probability of the privacy amplification procedure.

\nocite{*}
\bibliography{main}

\end{document}